\documentclass[10pt, a4]{article}
\usepackage[a4paper,bottom=1in,top=1in,left=1in,right=1in]{geometry}
\usepackage{graphicx}
\usepackage{amsmath}
\usepackage{amssymb}

\usepackage{epstopdf}
\usepackage{caption}
\usepackage{subcaption}
\usepackage{color}
\usepackage{comment}
\usepackage{soul}
\usepackage{tikz}
\usepackage{pgfplots}
\usepackage{tikz-3dplot}
\tdplotsetmaincoords{60}{115}
\pgfplotsset{compat=newest}
\usetikzlibrary{external}
\usepackage{lscape}

\usepackage{hyperref}
\hypersetup{
     colorlinks 		= true,
     linkcolor 		= blue,
     anchorcolor 		= blue,
     citecolor 		= blue,
     filecolor 		= blue,
     urlcolor 		= blue
     }

\newcommand*{\figref}[2][]{%
  \hyperref[{fig:#2}]{%
    \ref*{fig:#2}%
    \ifx\\#1\\%
    \else
      \,#1%
    \fi
  }%
}

\newcommand{\citet}[1]{\cite{#1}}

\begin{document}

\title{Revisiting boundary layer flows of viscoelastic fluids}

\author{L.~J.~Escott$^{1,}$\thanks{Email: \href{mailto:l.escott@aston.ac.uk}{l.escott@aston.ac.uk}}\,\,and P.~T.~Griffiths$^{1}$
\\
\\
$^{1}$Department of Mathematics, School of Informatics and Digital Engineering,
\\
Aston University, Birmingham B4 7ET, United Kingdom}

\date{}

\noindent This article has been accepted for publication in Journal of Non-Newtonian Fluid Mechanics. The DOI number for this manuscript is \href{https://doi.org/10.1016/j.jnnfm.2022.104976}{doi: 10.1016/j.jnnfm.2022.104976}

\vspace{\baselineskip}

\noindent Cite as: J.~Non-Newt.~Fluid~Mech. \textbf{312}, 104976 (2023)

\vspace{\baselineskip}

\begin{abstract}
In this article we reconsider high Reynolds number boundary layer flows of fluids with viscoelastic properties. We show that a number of previous studies that have attempted to address this problem are, in fact, incomplete. We correctly reformulate the problem and solve the governing equations using a Chebyshev collocation scheme. By analysing the decay of the solutions to the far-field we determine the correct stress boundary conditions required to solve problems of this form.

Our results show that both the fluid velocity within the boundary layer and the stress at the solid boundary increase due to the effect of viscoelasticity. As a consequence of this, we predict a thinning of the boundary layer as the value of the dimensionless viscoelastic flow parameter is increased. These results contradict a number of prominent studies in the literature but are supported by results owing from an asymptotic analysis based on the assumption of the smallness of the non-dimensional viscoelastic flow parameter.
\end{abstract}

\maketitle

\section{Introduction}\label{sec:intro}

Interest in the theory of viscoelastic boundary layer flows first developed in late 1950s and early 1960s, some 50 years, or so, after the fundamental contributions of \citet{Prandtl} and \citet{Blasius}. The early works of \citet{Srivastava}, \citet{Bhatnagar}, and \citet{RajeswariRathna} sought to develop the theory of, and approximate analytical solutions to, the stagnation point boundary layer flow of viscoelastic fluids described using Rivlin–Ericksen tensors~\cite{RivlinEricksen}.

Ultimately, each of these early studies suffered either from 
a misunderstanding of the governing physics and/or mathematical shortcomings. It wasn't until the slightly later study of \citet{BeardWalters} that the first complete, correct, boundary layer analysis concerning viscoelastic fluids was completed. In this study the authors begin by considering an Oldroyd-B constitutive viscosity law and then choose to restrict their attention to fluids with short relaxation times. This restriction reduces their analysis to consider what we would now refer to as a `second-order' fluid model. In the absence of advanced numerical techniques, the authors develop approximate analytical solutions using an asymptotic expansion based on their assumption of the smallness of the non-dimensional viscoelastic flow parameter. They conclude that as viscoelasticity is increased, so does the stress at the solid boundary and, subsequently, the boundary layer is thinned when compared to its Newtonian counterpart.

Following the work of \citet{BeardWalters} there have been a plethora of studies concerning boundary layer flows of viscoelastic fluids characterised by the second-order fluid model (sometimes referred to as `fluids of second grade'). Rajagopal and co-authors have produced numerous studies of this nature, see for example \citet{Rajagopaletal1980}, \citet{Rajagopaletal1983}, \citet{RajagopalGupta}, \citet{GargRajagopal1990}, \citet{GargRajagopal1991}. In these works the authors cover the theory of second-order fluid boundary layer flows, use the perturbation method introduced by \citet{BeardWalters} to analyse flows over wedges, investigate the effects of suction through a permeable surface, use the idea of an augmented boundary condition to arrive at new solutions to the stagnation point flow problem, and, lastly, show that their stagnation point analysis can be extended, in a local sense, to flows over wedges.

The use of Garg and Rajagopal's augmented boundary condition for problems of this nature has been discussed at some length in the literature. Indeed, we will subsequently be required to revisit these discussions as part of our analysis. At this stage, we refer the interested reader to the work of \citet{Ariel}, wherein the vast majority of suitable references on this discussion point can be found.

Seeking to improve upon the aforementioned second-order fluid analyses, in the mid 2000s, Sadeghy and co-authors focussed their efforts on the study of viscoelastic boundary layer flows with an upper-convected Maxwell (UCM) constitutive viscosity law (\citet{Sadeghyetal2005}, \citet{Sadeghyetal2006}). Given the relative limitations of the second-order fluid model, the motivation for these studies came from a want to capture the correct physics of these problems for larger values of the viscoelastic flow parameter. However, in both studies, the authors relied upon the problem formulation presented in Harris' book `\textit{Rheology and Non-Newtonian Flow}' (\citet{Harris}). Careful examination of this reference shows that Harris made a rather a significant oversight when deriving the equations relevant to UCM boundary layers. As such, the analysis presented by Sadeghy \emph{et al.}~fails to correctly capture the effects of viscoelasticity for high Reynolds number flows. This realisation is important, not only because of the need to produce mathematically correct solutions, but also because of the number of subsequent studies that have used this formulation type as a basis for their investigations.

To date, both \citet{Sadeghyetal2005} and \citet{Sadeghyetal2006} have received well over 100 citations each, with studies involving flows over porous surfaces (\citet{Hayatetal2006}), flows involving the transfer of mass (\citet{Hayatetal2011}), magnetohydrodynamic flows (\citet{Abeletal}) and rotating flows (\citet{Mustafaetal}), to name but a few, relying, via Sadeghy \emph{et al.}, on the mathematical shortcomings first presented by \citet{Harris}. Interestingly, we note that there is a far smaller readership for the work of \citet{PhanThien83}, published well in advance of both \citet{Sadeghyetal2005} and \citet{Sadeghyetal2006}, who use the correct equation set, contrary to those described by \citet{Harris}. This study conducted in the mid 1980s considers the stagnation point flow UCM problem, and is overlooked by Sadeghy and co-workers. The corresponding problem of the flow of an Oldroyd-B fluid is reported in \citet{PhanThien84}, and we note that a similar case study concerning the Giesekus fluid model is conducted by \citet{Mirzadeh2009}. Comparisons of our work with all these investigations is presented in \S \ref{sec:comp}.

Studies of boundary layer flows of other common viscoelastic models have been undertaken more recently. For example, \citet{Olagunju2006a} and \citet{Olagunju2006b} consider a FENE-P constitutive viscosity law, which seeks to improve on the Oldroyd-B model by introducing an elasticity which is finitely-extensible. These investigations focus on a forced convection boundary layer flow, and the calculation of local self-similar solutions for the flat plate boundary layer problem, respectively. We also note the work of \citet{Parvaretal2021}, who seek to extend the analysis presented by \citet{Olagunju2006b}. They combine semi-analytical results, in the sense of local self-similarity, with full numerical results owing from OpenFOAM simulations, showing a good agreement between the two sets of solutions. We go on to show in \S\ref{sec:form} that the flat plate geometry does not yield self-similar solutions for the Oldroyd-B model, and this is indeed the case for the FENE-P model, as was originally shown by \citet{Olagunju2006b}.

The motivation for this study stems from our work concerning the injection of non-Newtonian fluids into otherwise Newtonian boundary layer flows. In an effort to validate the results of our injection analyses, for flows involving viscoelastic fluids, we explored the large injection limit expecting our solutions to tend towards the known published results for entirely viscoelastic boundary layers. When this was not the case, we analysed the literature regarding flows of this nature and noted the mathematical shortcomings mentioned above. Before presenting the results of our much broader injection study we felt it important to revisit, and correct, the current state of the literature.

In this article, we have improved on the aforementioned studies in a number of ways. Firstly, where previously the necessity of a stagnation point flow has been assumed for flows of a viscoelastic fluid, we have worked independently of this assumption, and show analytically that one is restricted to this case study in order to obtain fully self-similar solutions. Secondly, we investigate the choice of stress boundary conditions in a full and rigorous manner, where in other articles the conditions are often simply stated or assumed without proper reference. In addition to this, we provide a detailed explanation of our numerical scheme, which we will show is highly accurate\footnote{Data created during this research can be found online at a location provided in~\ref{sec:SpectralData}}. Lastly, our solution method benefits from the possibility of evaluating moderate levels of viscoelastic effects. Previous studies \citet{Sadeghyetal2006}, \citet{PhanThien83}, \citet{PhanThien84}, \citet{Mirzadeh2009} only show results for a relatively low ratio of elastic to viscous effects, in some cases due to the imposition of a boundary condition which restricts the validity of the range of their dimensionless parameters.


The outline of this article is as follows, in \S\ref{sec:form} we formulate the problem, firstly in a general sense, and then derive the equations relevant to the flow of viscoelastic fluids described using the Oldroyd-B model, which similarly describes the UCM model by fixing a single parameter. In \S\ref{sec:res} we outline the numerical scheme used to solve the governing equations and present a range of results for fluids with varying levels of viscoelasticity. In \S\ref{sec:comp} we compare our results to those in the literature and discuss our findings in this context. Lastly, in \S\ref{sec:conc} we provide a brief summary and outline some future directions for work on problems such as this.

\section{Formulation}\label{sec:form}

Consider the steady flow of viscoelastic fluid over an impermeable, semi-infinite, flat plate inclined at an angle of $m\pi/(m+1)$, from the horizontal where $m$ is a constant that will be defined in due course. The streamwise coordinate is $x^{*}$ and the wall normal coordinate is $y^{*}$ (asterisks denote dimensional quantities throughout). The flow is governed by the continuity and Cauchy momentum equations
\begin{subequations}
\begin{align}
\boldsymbol{\nabla}^{*}\boldsymbol{\cdot}\boldsymbol{u}^{*}&=0,
\label{continuity}
\\
\rho^{*}(\boldsymbol{u}^{*}\boldsymbol{\cdot}\boldsymbol{\nabla}^{*}\boldsymbol{u}^{*})&=-\nabla^{*}p^{*}+\boldsymbol{\nabla}^{*}\boldsymbol{\cdot}\boldsymbol{\tau}^{*},
\label{cauchy}
\end{align}
\label{norm-gov}%
\end{subequations}
where $\rho^{*}$ is the fluid density, $p^{*}$ is the pressure, $\boldsymbol{u}^{*}=(u^{*},v^{*})$ is the two-dimensional velocity field with $u^{*}$ and $v^{*}$ being the streamwise and wall normal velocity components respectively. Lastly, we note that the definition of the stress tensor $\boldsymbol{\tau}^{*}$, changes according to the form of the viscoelastic model in question.

In this study we will focus our attention on three prominent viscoelastic models, these being the Oldroyd-B model, the upper-convected Maxwell model and the second-order fluid model. For the types of flows discussed in this article all, three models can be described using the following constitutive viscosity law
\begin{equation}
\boldsymbol{\tau}^{*}+\lambda_{1}^{*}\stackrel{\nabla^{*}}{\boldsymbol{\tau}^{*}}=2\mu_{0}^{*}(\boldsymbol{\mathrm{E}}^{*}+\lambda_{2}^{*}\stackrel{\nabla^{*}}{\boldsymbol{\mathrm{E}}^{*}}),\label{stress-gov}
\end{equation}
where $\mu_{0}^{*}$ is the total viscosity, $\boldsymbol{\mathrm{E}}^{*}=\left(\boldsymbol{\nabla}^{*}\boldsymbol{u}^{*}+(\boldsymbol{\nabla}^{*}\boldsymbol{u}^{*})^{\textrm{T}}\right)/2$, is the rate of strain tensor, $\lambda_{1}^{*}$ is the relaxation time, and $\lambda_{2}^{*}$ is the retardation time. We note that the velocity gradient is defined with the following convention in index notation: ${\left(\boldsymbol{\nabla}^{*}\boldsymbol{u}^{*}\right)}_{ij} = \boldsymbol{\nabla}^{*}_i\boldsymbol{u}^{*}_j$. The upper-convected derivative for a (steady) tensor $\boldsymbol{\mathrm{q}}^{*}$ is defined as follows
$$
\\
\stackrel{\nabla^{*}}{\boldsymbol{\mathrm{q}}^{*}}=\boldsymbol{u}^{*}\boldsymbol{\cdot}\boldsymbol{\nabla}^{*}\boldsymbol{\mathrm{q}}^{*}-(\boldsymbol{\nabla}^{*}\boldsymbol{u}^{*})^{\textrm{T}}\boldsymbol{\cdot}\boldsymbol{\mathrm{q}}^{*}-\boldsymbol{\mathrm{q}}^{*}\boldsymbol{\cdot}(\boldsymbol{\nabla}^{*}\boldsymbol{u}^{*}),
$$
which describes how a quantity is translated and rotated under the influence of the flow field. In the case when the total viscosity, relaxation time and retardation time are all non-zero, \eqref{stress-gov} returns the Oldroyd-B model. If the retardation time is set equal to zero then this model reduces to the upper-convected Maxwell model. In the limit of short relaxation times the equation of state \eqref{stress-gov} reduces to
$$\boldsymbol{\tau}^{*}=2\mu_{0}^{*}\boldsymbol{\mathrm{E}}^{*}-2\kappa_{0}^{*}\stackrel{\nabla^{*}}{\boldsymbol{\mathrm{E}}^{*}},$$
where $\kappa_{0}^{*}=\mu_{0}^{*}(\lambda_{1}^{*}-\lambda_{2}^{*})$ (see \citet{BeardWalters}). Now, the second-order fluid model is defined like so (see \citet{MorozovSpagnolie})
$$\boldsymbol{\tau}^{*}=2\mu_{0}^{*}\boldsymbol{\mathrm{E}}^{*}+2\alpha_{1}^{*}\stackrel{\nabla^{*}}{\boldsymbol{\mathrm{E}}^{*}}+4\alpha_{2}^{*}(\boldsymbol{\mathrm{E}}^{*}\cdot\boldsymbol{\mathrm{E}}^{*}).$$
While we consider the use of the upper-convected derivative, it is noted that other sources in the literature~\citet{KochSubramanian}, \citet{Rallison}, \citet{EscottWilson} use either the lower-convected or corotational derivative in order to describe the second-order fluid model. All definitions are equivalent so long as the material constants for the model are defined in an appropriate manner. Upon comparison of the definitions of the stress tensors above, we see that the limiting case of an Oldroyd-B fluid in short relaxation times does not capture the full general second-order fluid model~\cite{MorozovSpagnolie}, rather a specific case study. The simplest way of quantifying this is to set the constants $\alpha_{2}^{*}=0$, with $-\kappa_{0}^{*}=\alpha_{1}^{*}$. However, we show in~\ref{sec:SOFEEterm} that the full second-order fluid model is captured in our boundary layer analysis, since the quadratic rate of strain term, $\boldsymbol{\mathrm{E}}^{*}\cdot\boldsymbol{\mathrm{E}}^{*}$, results in no extra contribution to the terms associated with either the velocities or viscoelastic stresses. Rather, it provides a contribution to the isotropic pressure term. Therefore, the final set of boundary layer equations are independent of $\alpha_2^{*}$ under the relevant limit of time scales.

Here we solve the coupled system of equations (\eqref{norm-gov} -- \eqref{stress-gov}) by explicitly considering the independent stress components of $\boldsymbol{\tau}^{*}$ in two-dimensions. However, \citet{Harris} uses a separate and incomplete technique (in the UCM case study where $\lambda_2^{*}=0$), which involves the decoupling of velocity and stress. Firstly, Harris re-arranges equation~\eqref{cauchy} to define $\boldsymbol{\nabla}^{*}\boldsymbol{\cdot}\boldsymbol{\tau}^{*}$, the stress divergence, precisely in terms of the velocity and pressure. Then, the divergence operator is applied to~\eqref{stress-gov}, with the aim of capturing all stress dependent terms as a function of the same vector function, the stress divergence. One can see that, if this were possible, the definition of $\boldsymbol{\nabla}^{*}\boldsymbol{\cdot}\boldsymbol{\tau}^{*}$ could be inserted into this new equation such that we are left with one equation in velocity and pressure only. However, it is not possible to describe the viscosity law as such, since the upper-convected derivative and divergence are not commutative operators: $\boldsymbol{\nabla}^{*}\boldsymbol{\cdot}\stackrel{\nabla^{*}}{\boldsymbol{\mathrm{q}}^{*}}\,\not\equiv\,\stackrel{\nabla^{*}}{\boldsymbol{\mathrm{Q}}^{*}}$, where $\boldsymbol{\mathrm{Q}}^{*}=\boldsymbol{\nabla}^{*}\boldsymbol{\cdot}\boldsymbol{\mathrm{q}}^{*}$, for any general quantity $\boldsymbol{\mathrm{q}}^{*}$. We note that the vector equivalent of the (steady) upper-convected derivative is defined as $\stackrel{\nabla^{*}}{\boldsymbol{\mathrm{Q}}^{*}}=\boldsymbol{u}^{*}\boldsymbol{\cdot}\boldsymbol{\nabla}^{*}\boldsymbol{\mathrm{Q}}^{*}-(\boldsymbol{\nabla}^{*}\boldsymbol{u}^{*})\boldsymbol{\cdot}\boldsymbol{\mathrm{Q}}^{*}$.

To better visualise the inequivalence, we highlight a simple case study, where the quantity $\boldsymbol{\mathrm{q}}^{*}$ represents an isotropic tensor $\boldsymbol{\mathrm{I}}^{*}$. The upper-convected derivative of $\boldsymbol{\mathrm{I}}^{*}$ is some multiple of the rate of strain tensor; since the identity tensor is homogeneous it has no spacial or temporal dependence, and therefore the convective term depends entirely on the velocity gradient. It must also be the case that the upper-convected derivative is frame invariant, which reduces the quantity to the symmetric component of the velocity gradient i.e., the rate of strain tensor. After the explicit calculation, we confirm that this is the case, and further show that the divergence of $\stackrel{\nabla^{*}}{\boldsymbol{\mathrm{q}}^{*}}$ reduces to a velocity Laplacian
$$\boldsymbol{\nabla}^{*}\boldsymbol{\cdot}\stackrel{\nabla^{*}}{\boldsymbol{\mathrm{I}}^{*}}=\boldsymbol{\nabla}^{*}\boldsymbol{\cdot}\left[\boldsymbol{u}^{*}\boldsymbol{\cdot}\boldsymbol{\nabla}^{*}\boldsymbol{\mathrm{I}}^{*}-(\boldsymbol{\nabla}^{*}\boldsymbol{u}^{*})^{\textrm{T}}\boldsymbol{\cdot}\boldsymbol{\mathrm{I}}^{*}-\boldsymbol{\mathrm{I}}^{*}\boldsymbol{\cdot}(\boldsymbol{\nabla}^{*}\boldsymbol{u}^{*})\right]=-2\boldsymbol{\nabla}^{*}\boldsymbol{\cdot}\boldsymbol{\mathrm{E}}^{*}=-{\nabla^{*}}^2\boldsymbol{u}^{*}.$$
The other side of our non-commutative equation is represented by the upper-convected derivative of $\boldsymbol{\mathrm{Q}}^{*}$. From our definitions above clearly $\boldsymbol{\mathrm{Q}}^{*}=\boldsymbol{\nabla}^{*}\boldsymbol{\cdot}\boldsymbol{\mathrm{I}}^{*}=\boldsymbol{0}$, and thus the upper-convected derivative of this quantity must also be zero. This shows conclusively that the two operators are not commutative.

We start the correct analysis with governing equations \eqref{norm-gov} and \eqref{stress-gov}, which are made dimensionless via the introduction of the following variables
$$(x,y)=\frac{(\delta x^{*},y^{*})}{\delta L^{*}},\quad (u,v)=\frac{(\delta u^{*},v^{*})}{\delta U^{*}},\quad p=\frac{p^{*}}{\rho^{*}(U^{*})^{2}},\quad T_{i,j}=\frac{\tau_{i^{*},j^{*}}^{*}L^{*}}{\mu_{0}^{*}U^{*}},$$
where $\delta=(\rho^{*}U^{*}L^{*}/\mu_{0}^{*})^{-1/2}=\textrm{Re}^{-1/2}$, is the standard boundary layer length scale, and $L^{*}$ and $U^{*}$ are characteristic length and velocity scales, respectively. This choice of scaling leads to the definition of two viscoelastic dimensionless parameters
$$\textrm{Wi}=\frac{\lambda_{1}^{*}U^{*}}{L^{*}},\qquad\beta=\frac{\lambda_{1}^{*} - \lambda_{2}^{*}}{\lambda_{1}^{*}},$$
where $\textrm{Wi}$, the ratio of relaxation time and the flow time scale, is the Weissenberg number and $1 - \beta$ is the ratio of retardation and relaxation times respectively. We note that, under the definitions $\mu_{s}^{*} + \mu_{p}^{*} = \mu_{0}^{*}$ and $\lambda_{2}^{*}=\mu_s^{*} \lambda_{1}^{*} / \mu_0^{*} = \lambda_{1}^{*} ( 1 - \mu_p^{*} / \mu_{0}^{*} )$ with $\mu_s^{*}$ and $\mu_p^{*}$ the solvent and polymer viscosities respectively, $\beta$ also represents a ratio of viscosities, which must remain bounded: $0\leq\beta\leq1$.

In the first instance we will consider an Oldroyd-B (OB) analysis, this being the case when $\textrm{Wi}\neq0$, and $\beta\neq1$. In this instance, given the definitions of the dimensionless variables, it is relatively straightforward to show that the boundary layer flow is governed by the following system of equations
\begin{subequations}
\begin{align}
\partial_{x}u+\partial_{y}v&=0,
\label{con}
\\
u\partial_{x}u+v\partial_{y}u&=-\partial_{x}p+\delta^{2}\partial_{x}T_{xx}+\delta\partial_{y}T_{xy},
\label{momx}
\\
u\partial_{x}v+v\partial_{y}v&=-\delta^{-2}\partial_{y}p+\delta\partial_{x}T_{xy}+\partial_{y}T_{yy},
\label{momy}
\\
2\partial_{x}u&=T_{xx}+\textrm{Wi}\,[u\partial_{x}T_{xx}+v\partial_{y}T_{xx}-2(T_{xx}\partial_{x}u+\delta^{-1}T_{xy}\partial_{y}u)\nonumber
\\
&\quad+2(1-\beta)\delta^{-2}(\partial_{y}u)^{2}+4(1-\beta)(\partial_{x}u)^{2}+2(1-\beta)\partial_{y}u\partial_{x}v\nonumber \\
&\quad-2(1-\beta) u\partial_{xx}u-2(1-\beta) v\partial_{xy}u],\label{Txx}
\\
\delta^{-1}\partial_{y}u+\delta\partial_{x}v&=T_{xy}+\textrm{Wi}\,[u\partial_{x}T_{xy}+v\partial_{y}T_{xy}-(\delta^{-1}T_{yy}\partial_{y}u+\delta T_{xx}\partial_{x}v)\nonumber
\\
&\quad+3(1-\beta)(\delta^{-1}\partial_{y}u\partial_{y}v+\delta\partial_{x}u\partial_{x}v)-(1-\beta) v(\delta^{-1}\partial_{yy}u+\delta\partial_{xy}v)\nonumber
\\
&\quad+(1-\beta)(\delta^{-1}\partial_{x}u\partial_{y}u+\delta\partial_{x}v\partial_{y}v)-(1-\beta) u(\delta^{-1}\partial_{xy}u+\delta \partial_{xx}v)],\label{Txy}
\\
2\partial_{y}v&=T_{yy}+\textrm{Wi}\,[u\partial_{x}T_{yy}+v\partial_{y}T_{yy}-2(\delta T_{xy}\partial_{x}v+T_{yy}\partial_{y}v)\nonumber
\\
&\quad+2(1-\beta)\delta^{2}(\partial_{x}v)^{2}+4(1-\beta)(\partial_{y}v)^{2}+2(1-\beta)\partial_{y}u\partial_{x}v\nonumber \\
&\quad-2(1-\beta)v\partial_{yy}v-2(1-\beta) u\partial_{xy}v],\label{Tyy}
\end{align}
\end{subequations}
where the notation $\partial_{k}$ denotes the partial derivative with respect to $k$. At this stage of our investigation, we restrict the OB model to, at largest, moderate values of the Weissenberg number i.e., $\textrm{Wi}\sim\mathcal{O}(\delta^{0})$, and we note that different scalings are introduced if $\textrm{Wi}$ is assumed to
be large. All that remains now is to determine the relevant boundary layer scalings for the three stress components $(T_{xx},T_{xy},T_{yy})$. In the case when $\textrm{Wi}\to0$, we must return the standard Newtonian boundary layer equations. As such, it follows immediately that the correct scaling for the shear stress component is, $T_{xy}=\delta^{-1}\tau_{xy}$. Given this result, when considering \eqref{Tyy}, we see that in order to ensure the correct leading order balance it must follow that $T_{yy}=\delta^{0}\tau_{yy}$. Lastly, from \eqref{Txx}, we observe that a non-zero shear stress will only be predicted in the case when $T_{xx}=\delta^{-2}\tau_{xx}$.

Having now determined all the correct scales we replace $T_{ij}$ with $\tau_{ij}$ throughout \eqref{momx}--\eqref{Tyy}, and take the limit as $\textrm{Re}\to\infty$. We then arrive at the OB boundary layer equations 
\begin{subequations}
\begin{align}
\partial_{x}u+\partial_{y}v&=0,
\label{blcon}
\\
u\partial_{x}u+v\partial_{y}u&=-\textrm{d}_{x}p+\partial_{x}\tau_{xx}+\partial_{y}\tau_{xy},
\label{blmomx}
\\
0&=\tau_{xx}+\textrm{Wi}\,[u\partial_{x}\tau_{xx}+v\partial_{y}\tau_{xx}-2(\tau_{xx}\partial_{x}u+\tau_{xy}\partial_{y}u)+2(1-\beta)(\partial_{y}u)^{2}],
\label{tauxx}\\
\partial_{y}u&=\tau_{xy}+\textrm{Wi}\,\{u\partial_{x}\tau_{xy}+v\partial_{y}\tau_{xy}-(\tau_{yy}\partial_{y}u+\tau_{xx}\partial_{x}v)\nonumber
\\
&\quad+(1-\beta)[3\partial_{y}u\partial_{y}v-v\partial_{yy}u+\partial_{x}u\partial_{y}u-u\partial_{xy}u]\},
\label{tauxy}\\
2\partial_{y}v&=\tau_{yy}+\textrm{Wi}\,\{u\partial_{x}\tau_{yy}+v\partial_{y}\tau_{yy}-2(\tau_{xy}\partial_{x}v+\tau_{yy}\partial_{y}v)\nonumber
\\
&\quad+(1-\beta)[4(\partial_{y}v)^{2}-2v\partial_{yy}v+2\partial_{y}u\partial_{x}v-2u\partial_{xy}v]\}\label{tauyy},
\end{align}
\label{blgov}%
\end{subequations}
noting that to leading order the pressure is a function of $x$ only; this follows immediately from \eqref{momy}. In order to remove the pressure from the problem we consider the flow in the free-stream where the streamwise velocity is a function of $x$ only, and the wall normal velocity is identically zero. Outside of the boundary layer it must also be true that the stress varies only in the streamwise direction. Therefore, in the limit as $y\to\infty$, we have from \eqref{blmomx} that
$$u^{F}\textrm{d}_{x}u^{F}=-\textrm{d}_{x}p+\textrm{d}_{x}\tau_{xx}^{F},$$
where the superscript $F$ denotes a free-stream quantity. As a first attempt to solve system \eqref{blgov} we seek a self-similar solution of the form
$$u=\partial_{y}\Psi,\quad v=-\partial_{x}\Psi,\quad\Psi=\sqrt{xu^{F}}f(\eta),\quad\eta=y\sqrt{\frac{u^{F}}{x}},$$
where, at this stage, we do not specify any restrictions on the three stress components. From \eqref{blmomx} we have that
\begin{equation}
u^{F}\textrm{d}_{x}u^{F}\biggl[(f')^{2}-\frac{ff''}{2}-1\biggr]-\frac{(u^{F})^{2}ff''}{2x}=-\textrm{d}_{x}\tau_{xx}^{F}+\partial_{x}\tau_{xx}+\partial_{y}\tau_{xy},
\label{self_similar}
\end{equation}
where the primes indicate differentiation with respect to the similarity variable $\eta$. In order to achieve a self-similar solution, we must guarantee that all the terms in above equation have the same $x$ dependence. Given the form of the left-hand side of the \eqref{self_similar}, this concisely forces the definition of the free-stream velocity, such that, $u^{F}=x^{m}$, where $m$ dictates the angle of inclination of the flat plate as noted at the outset of this section. In turn, this implies that $\tau_{xy}=x^{(3m-1)/2}s_{xy}(\eta)$, and that $\tau_{xx}=x^{2m}s_{xx}(\eta)$. Therefore, upon inserting these three functions of $x$ into \eqref{self_similar}, we arrive at
$$m[(f')^{2}-1]-\frac{(m+1)ff''}{2}=-x^{1-2m}\textrm{d}_{x}\tau_{xx}^{F}+2ms_{xx}+\frac{(m-1)\eta s_{xx}'}{2}+s_{xy}'.$$
It must then follow that $\tau_{xx}^{F}=Cx^{2m}$, where the constant $C$ is determined from the solution of \eqref{tauxx} in the limit as $y\to\infty$. One finds that, in this limit, a general solution of \eqref{tauxx} can only be found in the instance when $C$ is identically zero and we must then have that $s_{xx}\to0$, as $\eta\to\infty$.

\begin{figure}[!t]
\centering
\includegraphics[width=.75\textwidth]{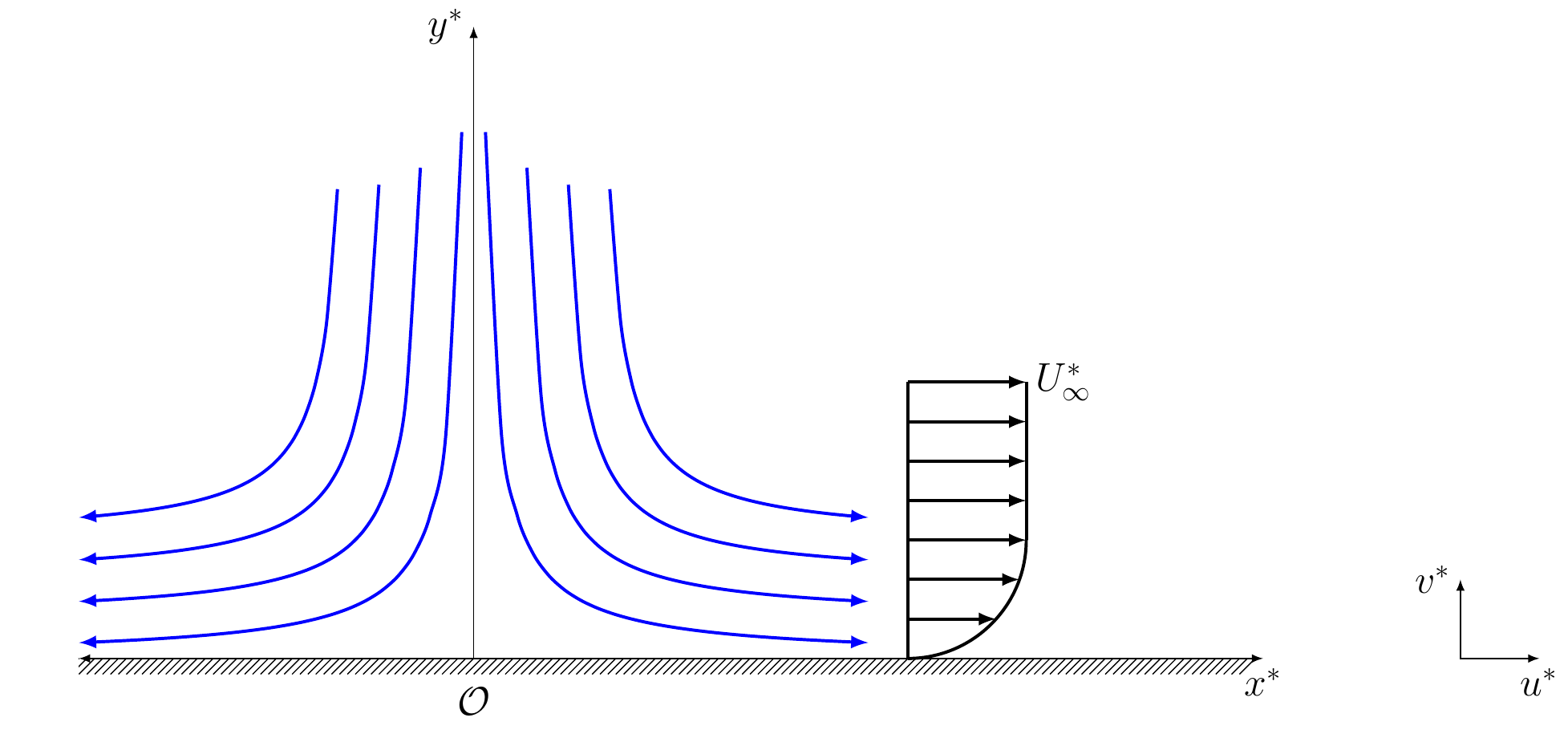}
\caption{Schematic diagram of the stagnation point boundary layer flow over a wedge inclined at angle of $\pi/2$ from the horizontal. Sketches of the streamlines are indicated in blue. A typical boundary layer velocity profile with free-stream velocity, $U_{\infty}^{*}$, is highlighted in black. All stated quantities are dimensional.}
\label{fig:schematic}
\end{figure}

In order to determine the range of values of $m$ for which self-similar solutions can exist, we must investigate the form of the governing stress equations. Analysis of \eqref{tauxx}--\eqref{tauyy} reveals that $\tau_{yy}=x^{m-1}s_{yy}(\eta)$, and that $x$ independence can only be ensured in the specific case when $m=1$, i.e., when one considers stagnation point flow over a wedge inclined at an angle of $\pi/2$ from the horizontal. We provide a schematic diagram of this flow configuration in Figure \ref{fig:schematic}. In this specific case we have that $u=xf'(y)$, $v=-f(y)$, $\tau_{xx}=x^{2}s_{xx}(y)$, $\tau_{xy}=xs_{xy}(y)$, and that $\tau_{yy}=s_{yy}(y)$. We note that primes now indicate differentiation with respect to $y$.

In order to simplify the forthcoming analysis it proves useful to introduce the following transformations $\mathcal{T}_{xx}=s_{xx}$, $\mathcal{T}_{xy}=s_{xy}-f''$, and $\mathcal{T}_{yy}=s_{yy}+2f'$. Having done so, we are able to express the problem as a system of six coupled first order ordinary differential equations


\begin{subequations}
\begin{align}
f'&=g,
\quad g'=h,
\quad
h'=g^{2}-1-fh-2\mathcal{T}_{xx}-\mathcal{T}_{xy}',\tag{6a, 6b, 6c}
\\
\mathcal{T}_{xx}&=\textrm{Wi}\,[2h(\beta h+\mathcal{T}_{xy})+f\mathcal{T}_{xx}'],\tag{6d}
\\
\mathcal{T}_{xy}&=\textrm{Wi}\,[f(\beta h'+\mathcal{T}_{xy}')-g(3\beta h+\mathcal{T}_{xy})+h\mathcal{T}_{yy}],\tag{6e}
\\
\mathcal{T}_{yy}&=\textrm{Wi}\,[2g(2\beta g-\mathcal{T}_{yy})-f(2\beta h-\mathcal{T}_{yy}')].\tag{6f}
\end{align}
\label{govODEs}%
\end{subequations}
This system must be closed subject to six boundary conditions. Given that the surface of the plate is impermeable we must have that $v(y=0)=0$, thus $f(y=0)=0$. The no-slip condition at the surface implies that $u(y=0)=0$, and hence $g(y=0)=0$. Given that the boundary layer flow must match with the free-stream above, we have the condition $u(y\to\infty)\to u^{F}$, which implies that $g(y\to\infty)\to1$. Our final three conditions must relate to the three stress components. 

We have already shown that $\mathcal{T}_{xx}(y\to\infty)\to0$, it remains, therefore, for us to arrive at physically relevant conditions for the functions $\mathcal{T}_{xy}$ and $\mathcal{T}_{yy}$. We note that as we approach the limit of infinite distance from the wall, both $\tau_{xy}$ and $\tau_{yy}$ must tend to functions of $x$, the variable of flow direction, and be independent of $y$. This is not to say that the functions must be zero, in fact we will go on to show that this is not the case. However, we must have no change in the stress contributions with respect to $y$, otherwise there will be a discontinuity in our solutions at the free-stream. It follows that our final two boundary conditions are then $s'_{xy}(y\to\infty)\to0$, and $s'_{yy}(y\to\infty)\to0$. In summary, we must solve system \eqref{govODEs} with respect to the following six boundary conditions
\begin{align*}
&f(y=0)=g(y=0)=0,\quad g(y\to\infty)\to1,\\
&\mathcal{T}_{xx}(y\to\infty)\to0,\quad\mathcal{T}_{xy}'(y\to\infty)\to-h'_{\infty},\quad\mathcal{T}_{yy}'(y\to\infty)\to2h_{\infty},
\end{align*}
where the infinity subscript denotes evaluation in the limit as $y\to\infty$. In order to determine the values for the constants $h_{\infty}$, and $h_{\infty}'$, we look to the decay of the solutions in the far-field.

Given that the streamwise velocity function $g$ tends to unity as $y\to\infty$, then $f\to (y-\delta_{1})$, where
$$\delta_{1}=\int_{0}^{\infty}(1-f')\,\textrm{d}y=y_{\infty}-f_{\infty},$$
is the displacement thickness. It proves useful, in what follows, to introduce the shifted coordinate $Y=y-\delta_{1}$. Given the far-field conditions on $\mathcal{T}_{xx}$, and $\mathcal{T}_{xy}'$, we have from \textcolor{blue}{(6c)} that the function $h$ must satisfy
$$h'-h'_{\infty}=-Yh,$$
in the limit as $y\to\infty$. This ODE has the solution
$$h\to c\,\textrm{e}^{-(Y^{2}-\delta_{1}^{2})/2}+\sqrt{2}h'_{\infty}D\biggl(\frac{Y}{\sqrt{2}}\biggr),$$
where $c$ is a constant of integration and $D$ is Dawson's Integral function. Then, from \textcolor{blue}{(6f)} it immediately follows that
$$\mathcal{T}_{yy}\to\frac{4\,\textrm{Wi}}{1+2\,\textrm{Wi}}\bigg[\beta+\frac{Y(h_{\infty}-\beta h)}{2}\biggr],$$
as $y\to\infty$. In order to ensure matching with the free-stream it must then be true that $h_{\infty}=0$: if this is not the case then the solution for $\mathcal{T}_{yy}$ grows linearly with respect to $y$ at the outer edge of the boundary layer. Given this form for the solution for $\mathcal{T}_{yy}$, in the limit as $y\to\infty$, from \textcolor{blue}{(6e)}, one can show that
$$\mathcal{T}_{xy}\to-a_{1}\beta h(3-2a_{2}+a_{2} Yh+Y^{2})-a_{1}(1-\beta) h_{\infty}'Y,$$
where
$$a_{n}=\frac{n\textrm{Wi}}{(1+n\textrm{Wi})}.
$$
Again, in order to ensure matching with the free-stream it must then be true that $h_{\infty}'=0$, if this is not the case then the absolute value of the solution for $\mathcal{T}_{xy}$ grows linearly with respect to $y$ at the outer edge of the boundary layer (note that this conclusion could also have been drawn from analysis of the expression for $\mathcal{T}_{yy}$). Thus, the function $h$ decays to zero exponentially,  $h\to c\,\textrm{e}^{-(Y^{2}-\delta_{1}^{2})/2}$, irrespective of the value of the Weissenberg number. Therefore, as $y\to\infty$,  the three stress functions decay to their respective constant values as follows
\begin{subequations}
\begin{align}
    \mathcal{T}_{xx}&\to a_{1}\beta h^{2}[1-2a_{2}h(3Y)^{-1}+(a_{1}^{-1}+2a_{2}-3)Y^{-2}]\sim\textrm{e}^{-y^{2}},
    \\
    \mathcal{T}_{xy}&\to a_{1}\beta h[a_{2}(2+h')-(3+Y^{2})]\sim y^{2}\textrm{e}^{-y^{2}/2},
    \\
    \mathcal{T}_{yy}&\to a_{2}\beta(2+h')\sim 2a_{2}\beta+y\textrm{e}^{-y^{2}/2}.
\end{align}
\label{decay}%
\end{subequations}
The result for the function $\mathcal{T}_{xx}$ follows directly from integrating \textcolor{blue}{(6d)} with the the known large-$y$ expression for $\mathcal{T}_{xy}$. This ODE has a decaying solution that can be expressed in terms of the function $h$, the upper incomplete gamma function and the complementary error function. To leading order, the terms involving upper incomplete gamma function and the complementary error function can be approximated in terms of powers and $h$ and $Y$, and the result above follows. In order to ensure some level of brevity, the details of this calculation are included for the interested reader in the \S\ref{sec:append}.

Given the above analysis, we are now in a position to solve the system of six coupled first order ODEs that govern OB stagnation point flow, \eqref{govODEs}, subject to the six physically correct boundary conditions

\begin{subequations}
\begin{align}
&f(y=0)=g(y=0)=0,\quad g(y\to\infty)\to1,\label{govBCsa}\\
&\mathcal{T}_{xx}(y\to\infty)\to0,\quad\mathcal{T}_{xy}'(y\to\infty)\to0,\quad\mathcal{T}_{yy}'(y\to\infty)\to0.
\end{align}
\label{govBCs}%
\end{subequations}

\section{Numerical Method and Results}\label{sec:res}

We solve \eqref{govODEs} subject to \eqref{govBCs} using a spectral method, in particular a variation on the Chebyshev collocation scheme. We split the one dimensional domain, $y=[0,y_{\infty}]$, into $N + 1$ points, including the surface of the flat plate and the far-field location, which we define to be at $y_0 = 0$ and $y_{N} = y_{\infty}$, respectively. At this stage, we choose not to associate a specific value with $y_{\infty}$, instead we keep the formulation general and determine a value for $y_{\infty}$ based on the outcome of our numerical testing procedure. The solution points are linearly related to the Chebyshev collocation points $\chi_n = \textrm{cos} \left( \pi n / N \right)$, with $n = 0, 1, ... , N$, such that $y_n = y_{\infty} \left( 1 - \chi_n \right) / 2$, covers the entire domain. The majority of the interesting dynamics associated with this problem occur in a close proximity to the flat plate, which can be better approximated by these collocation points as opposed to a uniform grid, given the higher concentration of points at both ends of the $y$-domain.

We approximate all the flow quantities, $q$, as a finite sum
$$q(y) = \sum_{j=0}^{N}{}'' a_j T_j^{*}(y),$$
where ${}''$ denotes that the first and last terms of the sum are halved, $a_j$ represents the set of constants to be fitted from our discrete solutions, and $T_j^*$ is related to the $j$th Chebyshev polynomial of the first kind $T_j$, applied at location $\chi_n$, via
$$T_j^*(y_n) = T_j ( 1 - 2 y_n / y_{\infty} ) = T_j(\chi_n).$$
This form for the yet unknown functions is applied to all the quantities present in both the governing equations~\eqref{govODEs}, and boundary conditions~\eqref{govBCs}, including the stress components and their derivatives. The one exemption being $h'$, which is obtained post-hoc. The polynomials themselves then form a set of orthonormal basis functions which are widely used in interpolation and optimisation problems, not least due to their property of having a maximum magnitude of $1$ in the range $\chi \in [-1, 1]$. One can define these functions using their orthogonality condition with respect to the inner product in integral form, but a simpler definition follows directly from the recurrence relation
$$T_0(\chi) = 1, \qquad \qquad T_1(\chi) = \chi, \qquad \qquad T_j(\chi) = 2 \chi T_{j-1}(\chi) - T_{j-2}(\chi).$$

Using the discrete orthogonality relation
$$\sum_{n=0}^{N}{ }'' \, T_i^*(y_n) T_j^*(y_n) = \left\{
\begin{array}{lr}
0, & i \neq j\\
N / 2, & i = j \neq 0, N\\
N, & i = j = 0, N
\end{array}
\right.,$$
we can then calculate the unknown constants
$$a_j = \frac{2}{N} \sum_{n=0}^{N}{}'' q(y_n) T_j^*(y_n),$$
where $q(y_n)$ is the value of a flow quantity at a point in the discrete domain. These definitions allow us to return a continuous function for any quantity $q$, which is known exactly at the $N + 1$ distinct points. Furthermore, we may find the derivative of a generic quantity, $q'$, which is described by way of a differentiation matrix
$$A_{ij} = \frac{2 b_j}{N} \sum_{n=0}^{N}{}'' \, T_n^{*'}(y_i) T_n^*(y_j), \qquad \qquad b_j = \left\{
    \begin{array}{lr}
    1/2, & j = 0, N\\
    1, & j \neq 0, N
    \end{array}
    \right.,$$
which satisfies $q'(y_i) = \sum_{j=0}^N A_{ij} q(y_j)$, where $T_n^{*'}$ represents the derivative of Chebyshev polynomial $T_n^*$ with respect to $y$. This statement holds true for general choice of flow quantity, which is particularly useful when $q$ itself is a derivative.

The governing equations~\eqref{govODEs} are converted into a set of 9 closed equations, taking care to account for the number of boundary conditions noted in~\eqref{govBCs}. The first 4 equations represent the conservation of momentum and constitutive stress relations. These are implemented at every point in the $y$-domain, which therefore provides $4 N + 4$ equations to solve. Where a flow quantity, for example, $g$, is known at either boundary due to our imposed conditions, this value is implemented directly in the numerical scheme and is not solved for. We also note that the only function that is not directly determined as part of our numerical scheme is $h'$. Therefore, it is necessary to make explicit use of the differentiation matrix $A_{ij}$, in the determination of this function only.

The remaining governing equations describe the first order derivatives of the 5 primary flow functions $f,g,\mathcal{T}_{xx},\mathcal{T}_{xy},$ and $\mathcal{T}_{yy}$. We do so by making use of the differentiation matrix as described overleaf, solving, respectively, for
\begin{alignat*}{2}
    g(y_i)\left(1-\delta_{iN}\right) &=  A_{i0} f(y=0) + \sum_{j=1}^N A_{ij} f(y_j) - \delta_{iN} g(y=y_\infty), \quad \quad \quad &&i = 1, ..., N \\
    h(y_i) &= A_{i0} \, g(y=0) + A_{iN} \, g(y=y_{\infty}) + \sum_{j=1}^{N-1} A_{ij} g(y_j), \quad \quad \quad &&i = 1, ..., N - 1 \\
    \mathcal{T}_{xx}'(y_i) &= A_{iN} \mathcal{T}_{xx}(y=y_{\infty}) + \sum_{j=0}^{N-1} A_{ij} \mathcal{T}_{xx}(y_j), \quad \quad \quad &&i = 0, ..., N - 1. \\ \mathcal{T}_{xy}'(y_i) &= \sum_{j=0}^N A_{ij} \mathcal{T}_{xy}(y_j), \quad \quad \quad &&i = 0, ..., N-1. \\
    \mathcal{T}_{yy}'(y_i) &= \sum_{j=0}^N A_{ij} \mathcal{T}_{yy}(y_j), \quad \quad \quad &&i = 0, ..., N-1,
\end{alignat*}
where $\delta_{iN}$ represents the Kronecker delta. The $y_i$ locations at which we solve the first order system above have been chosen to remove degrees of freedom equal to the number of boundary conditions, such that we have a total of $9 N + 3$ equations to solve numerically. We choose to remove either one or two equations from the system based on whether there is a fixed boundary condition on a particular quantity or its integral. The first study involves a relation where the quantity summed over has at least one boundary condition attached to it. Take, for example, the function $h(y_i)$, as described above. We evaluate this function at all solution points apart from at the flat plate ($y=0$) and the far-field location ($y=y_{\infty}$), since we constrain its integral $g(y_i)$ at these points.

The other case is considered in, for example, the calculation of $\mathcal{T}_{xy}'(y_i)$, where the boundary condition is applied to the quantity on the left-hand side. In this case, since no boundary conditions are placed on $\mathcal{T}_{xy}$, we are required to solve for it at all points $y_n$. However, since we know the value of $\mathcal{T}_{xy}'(y=y_{\infty})$, the location $i=N$ is excluded from the calculations. This is equivalent to the statement that we know the value of $q'$ at a certain location, so we should not solve for it. We note however that this is not true of the first equation for $g(y_i)$ above, which does constrain $f$ with the known condition $g(y_\infty)\to1$. We justify this by recognising that $f$ does have an attributed boundary condition, and so falls into the first category of equation removals described above.

We use MATLAB to solve the system as described above, with the Levenberg-Marquardt algorithm up to a norm mean square error of $10^{-18}$. The numerical solutions presented here are calculated with a value of $N + 1 = 200$ solution points for a variety of both $\textrm{Wi}$ and $\beta$ values. 

In order to determine a suitable value for the location of the far-field boundary $y_{\infty}$, we solved the Newtonian flow problem ($\textrm{Wi}=0$) for a range of combinations of both $N$ and $y_{\infty}$. In each case we maintain the norm mean square error tolerance of $10^{-18}$. These Newtonian results were also compared to solutions obtained from a fourth-order Runge-Kutta integration scheme twinned with a secant shooting method. The difference between the results obtained via these schemes was numerically indistinguishable. Moving forward, we set $y_{\infty}=5$ to be the free-stream location, which captures the dynamics in the far-field in precisely the same way as any larger value for each of our viscoelastic models.

In the first instance, we will consider solutions to \eqref{govODEs} subject to \eqref{govBCs} for an arbitrary fixed value of the dimensionless retardation parameter, $\beta=0.8$. We note that the special case when $\beta=1$, corresponding to a system with an upper-convected Maxwell constitutive viscosity law, will be covered, in detail, in \S\ref{sec:comp}. 

\begin{figure}[t!]
\centering
\begin{subfigure}[b]{0.48\textwidth}
\centering
\includegraphics[width=\textwidth]{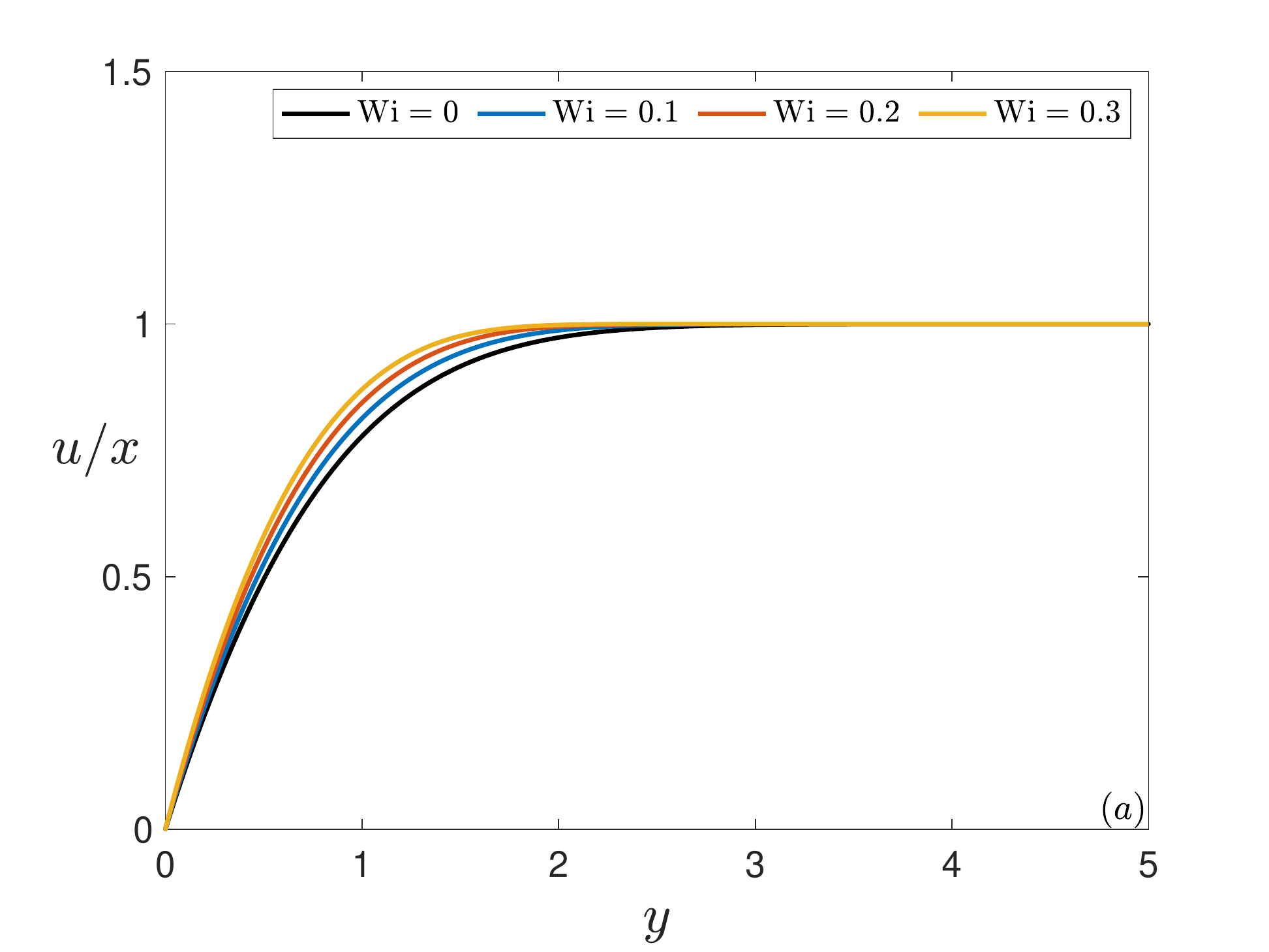}
\label{fig:beta_var_Wi_u}
\end{subfigure}
\begin{subfigure}[b]{0.48\textwidth}
\centering
\includegraphics[width=\textwidth]{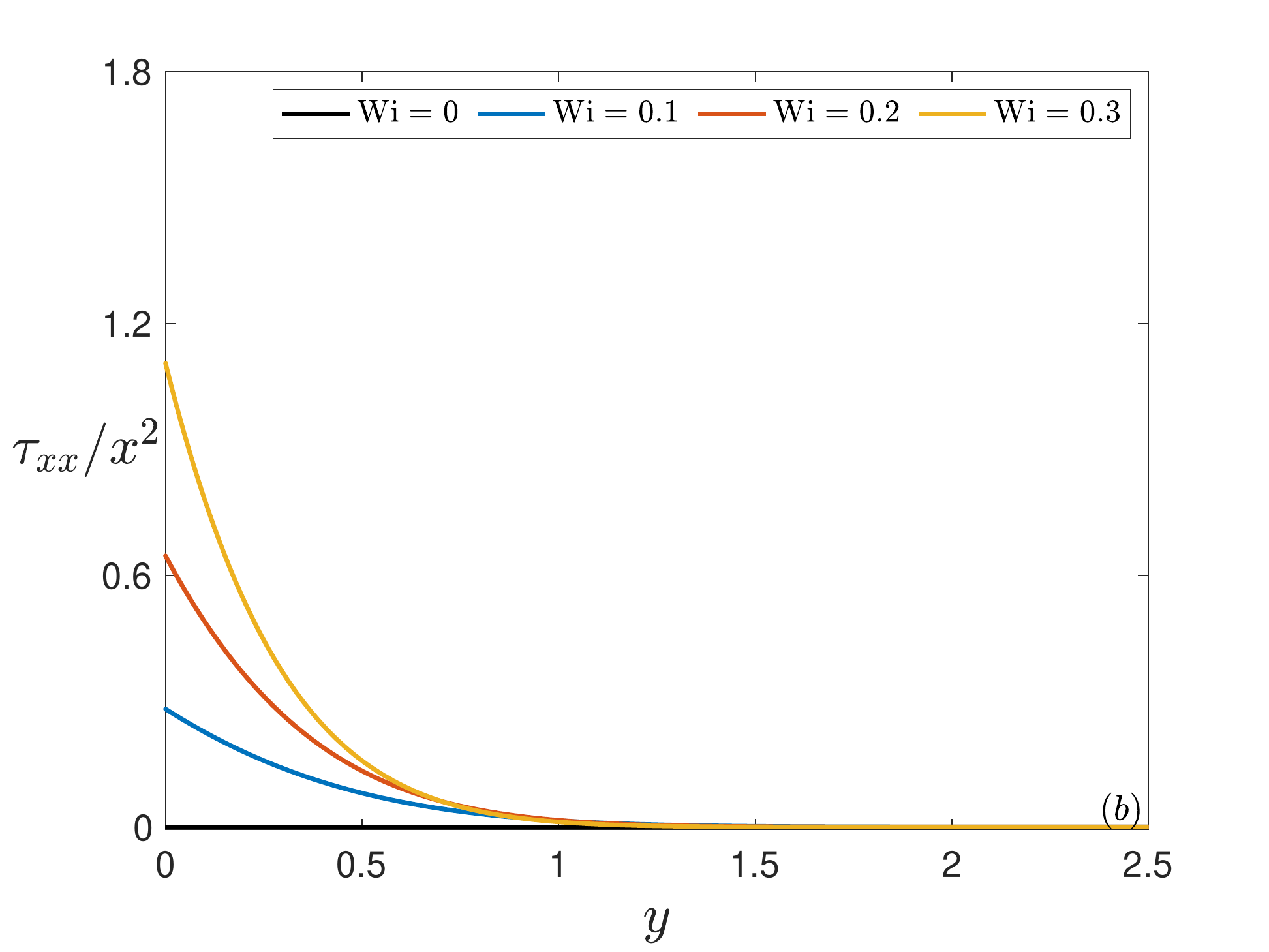}
\label{fig:beta_var_Wi_tau_xx}
\end{subfigure}
\begin{subfigure}[b]{0.48\textwidth}
\centering
\includegraphics[width=\textwidth]{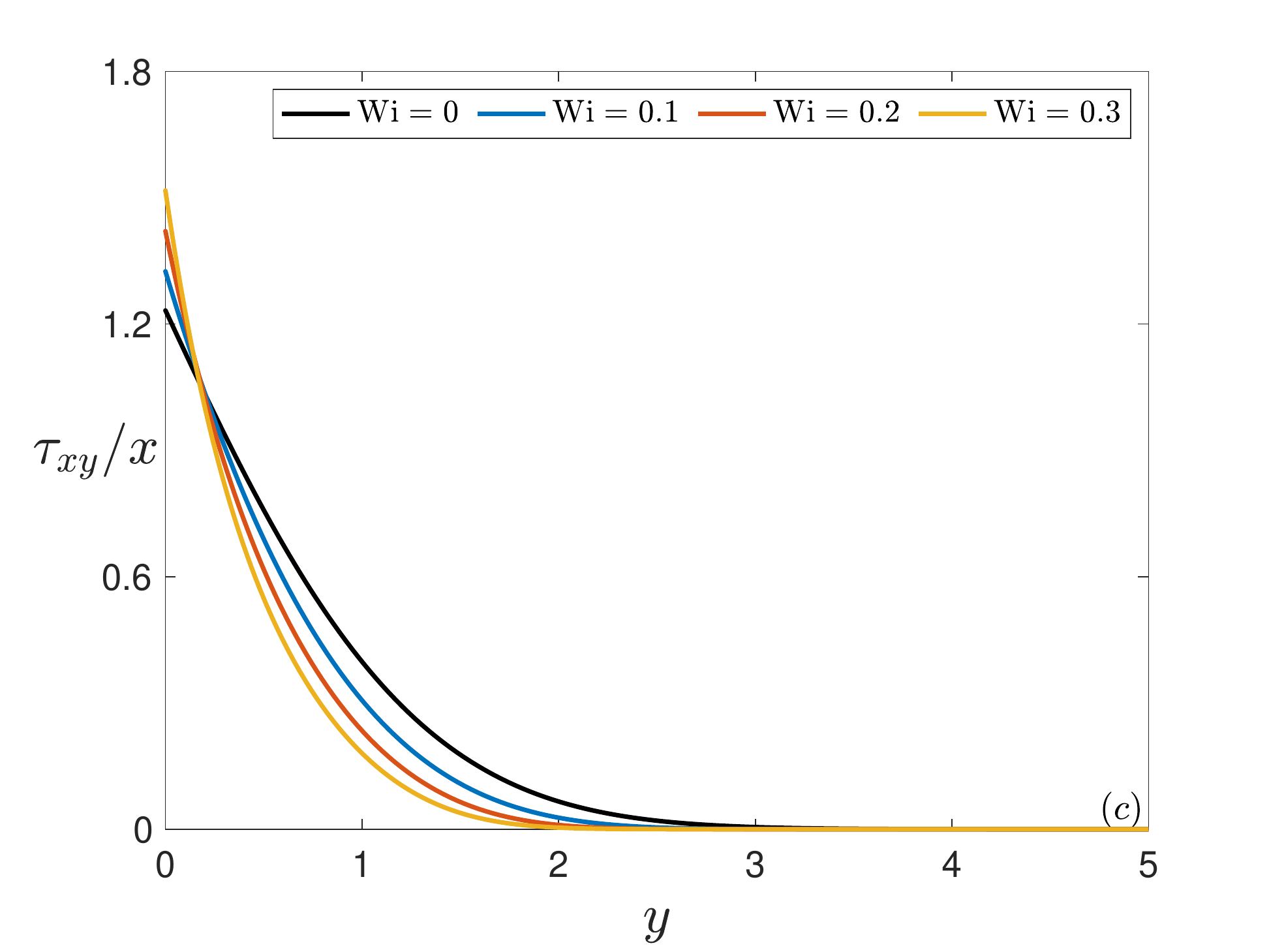}
\label{fig:beta_var_Wi_tau_xy}
\end{subfigure}
\begin{subfigure}[b]{0.48\textwidth}
\centering
\includegraphics[width=\textwidth]{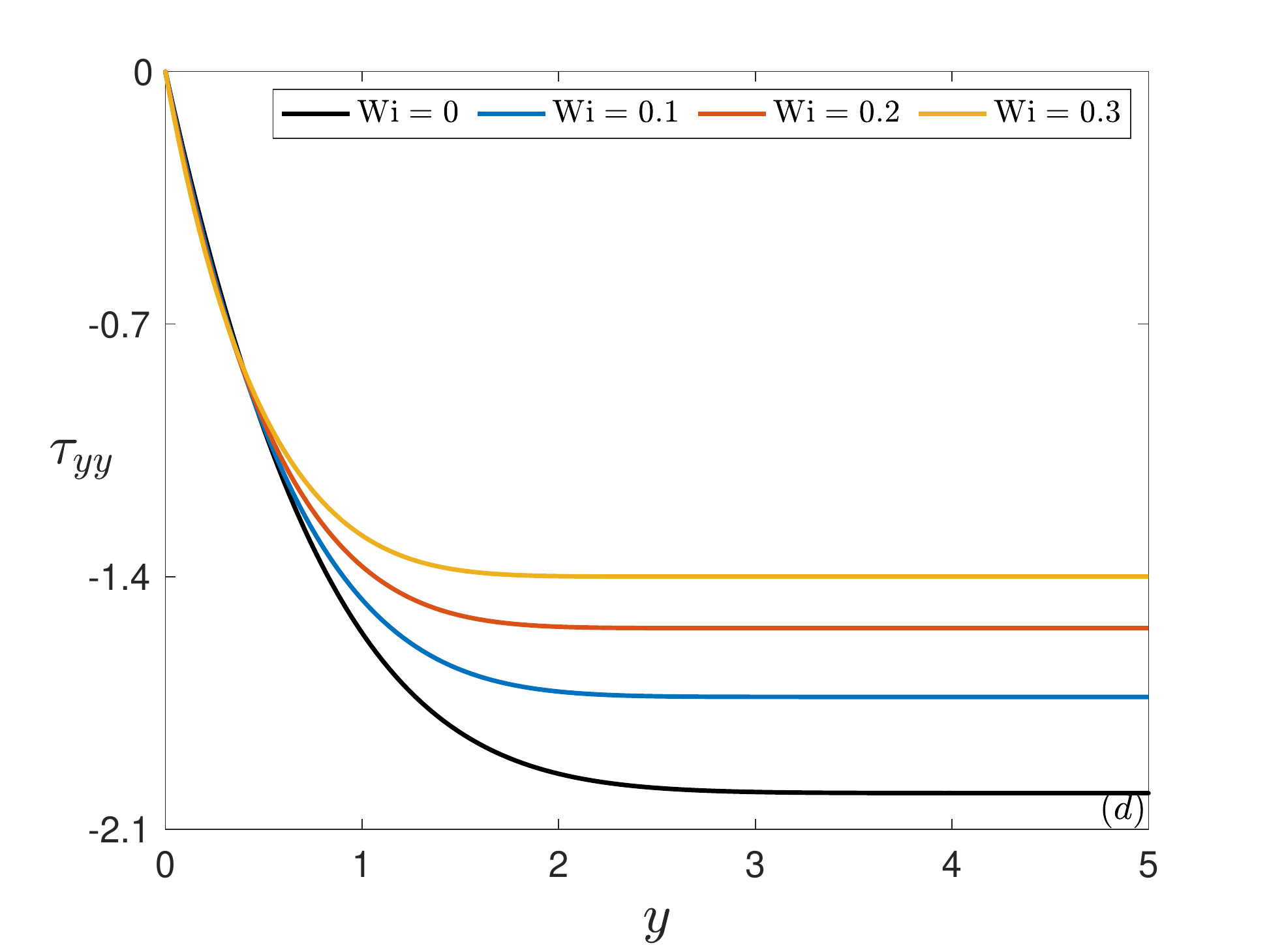}
\label{fig:beta_var_Wi_tau_yy}
\end{subfigure}
\caption{In (a) we plot the streamwise velocity $u/x$, against $y$, for a range of small values of the Weissenberg number. In (b), (c), and (d) we plot the variation of the three stress functions for the same range of values of the Weissenberg number. In all cases the dimensionless retardation parameter is fixed such that $\beta=0.8$.}\label{fig:beta_var_Wi}
\end{figure}

Given the the initial conditions noted in \eqref{govBCsa}, from \textcolor{blue}{(6f)} it is clear that $\mathcal{T}_{yy}(0)=s_{yy}(0)=\tau_{yy}(0)=0$. It then follows from \textcolor{blue}{(6e)}, that $\mathcal{T}_{xy}(0)=0$, thus $s_{xy}(0)=x^{-1}\tau_{xy}(0)=h(0)$. Lastly, from \textcolor{blue}{(6d)}, we must have that $\mathcal{T}_{xx}(0)=s_{xx}(0)=x^{-2}\tau_{xx}(0)=2\beta\textrm{Wi}\,[h(0)]^{2}$. Therefore, for any fixed value of $\beta$, there will exist a value of the Weissenberg number, $\textrm{Wi}=[2\beta h(0)]^{-1}$, such that $h(0)=s_{xx}(0)=s_{xy}(0)$. The value of these functions at the wall are closely related to two important parameters for flat plate viscoelastic boundary layer flows; the skin friction coefficient and the first normal stress difference. Firstly, the skin friction coefficient is defined as follows:
\begin{align*}
C_{f}=\biggl[\frac{2\tau_{x^{*}y^{*}}^{*}}{\rho^{*}(U^{*})^{2}}\biggr]\biggl|_{y^{*}=0}&=2\textrm{Re}^{-1/2}(\tau_{xy})|_{y=0}
\\
&=2x\,\textrm{Re}^{-1/2}h(0)
\\
&\quad-2x\,\textrm{Wi}\,\textrm{Re}^{-1/2}[g(0)s_{xy}(0)-f(0)s_{xy}'(0)-h(0)s_{yy}(0)]
\\
&\quad+2(1-\beta) x\,\textrm{Wi}\,\textrm{Re}^{-1/2}[3g(0)h(0)-f(0)h'(0)].
\end{align*}
Given that $f(0)=g(0)=s_{yy}(0)=0$, neither $\textrm{Wi}$, nor $\beta$ appear explicitly in the expression for the Oldroyd-B skin friction coefficient. Therefore, $C_{f}=2x\,\textrm{Re}^{-1/2}h(0)$, which is identical to the standard Newtonian expression. The other important quantity for flows such as these is the first normal stress difference at the wall:
$$
N_{1}=\biggl[\frac{2(\tau_{x^{*}x^{*}}^{*}-\tau_{y^{*}y^{*}}^{*})}{\rho^{*}(U^{*})^{2}}\biggr]\biggl|_{y^{*}=0}=2(\tau_{xx}-\textrm{Re}^{-1}\tau_{yy})|_{y=0}
=4\beta x^{2}\,\textrm{Wi}\,[h(0)]^{2}.
$$
We observe that the first normal stress difference at the wall is directly proportional to the square of the skin friction coefficient, $N_{1}=\beta\textrm{Wi}\,\textrm{Re}\,C_{f}^{2}$.

\subsection{Small \texorpdfstring{$\textnormal{Wi} \sim O(1)$}{}}\label{subsec:smallWiRes}

In the first instance we consider low Weissenberg number flows, which is in keeping with previous analyses available in literature. A comparison between our solutions and those of preceding studies can be found in \S \ref{sec:comp}.

\begin{table}
\begin{center}
\begin{tabular}{|c|c|c|c|c|}
\hline
$\textrm{Wi}$ & $x^{-1}\textrm{Re}^{1/2}C_{f}$ & $x^{-2}N_{1}$ & $(\tau_{yy})_{\infty}$ & $\delta_{1}$ \\
 & $=2h(0)$ & $=4\beta\textrm{Wi}\,[h(0)]^{2}$ &  &  \\
\hline
$0\hphantom{.00000}$        & $2.4652$ & $0\hphantom{.0000}$    & $-2\hphantom{.0000}$  & $0.6479$\\
$0.1\hphantom{0000}$        & $2.6514$ & $0.5624$               & $-1.7333$             & $0.5893$\\
$0.2\hphantom{0000}$        & $2.8430$ & $1.2932$               & $-1.5429$             & $0.5429$\\
$0.3\hphantom{0000}$        & $3.0361$ & $2.2123$               & $-1.4\hphantom{000}$  & $0.5058$\\
\hline
\end{tabular}
\caption{Base flow data for a range of values of the Weissenberg number in the case when $\beta=0.8$. We note that, in the limit as $y\to\infty$, the value of wall normal stress function, $(\tau_{yy})_{\infty}$, does not need to be computed numerically, instead, it is calculated using the asymptotic result that $(\tau_{yy})_{\infty}\to4\beta\textrm{Wi}/(1+2\textrm{Wi})-2$.}
\label{bfs}
\end{center}
\end{table}

\begin{figure}
\centering
\includegraphics[width=.75\textwidth]{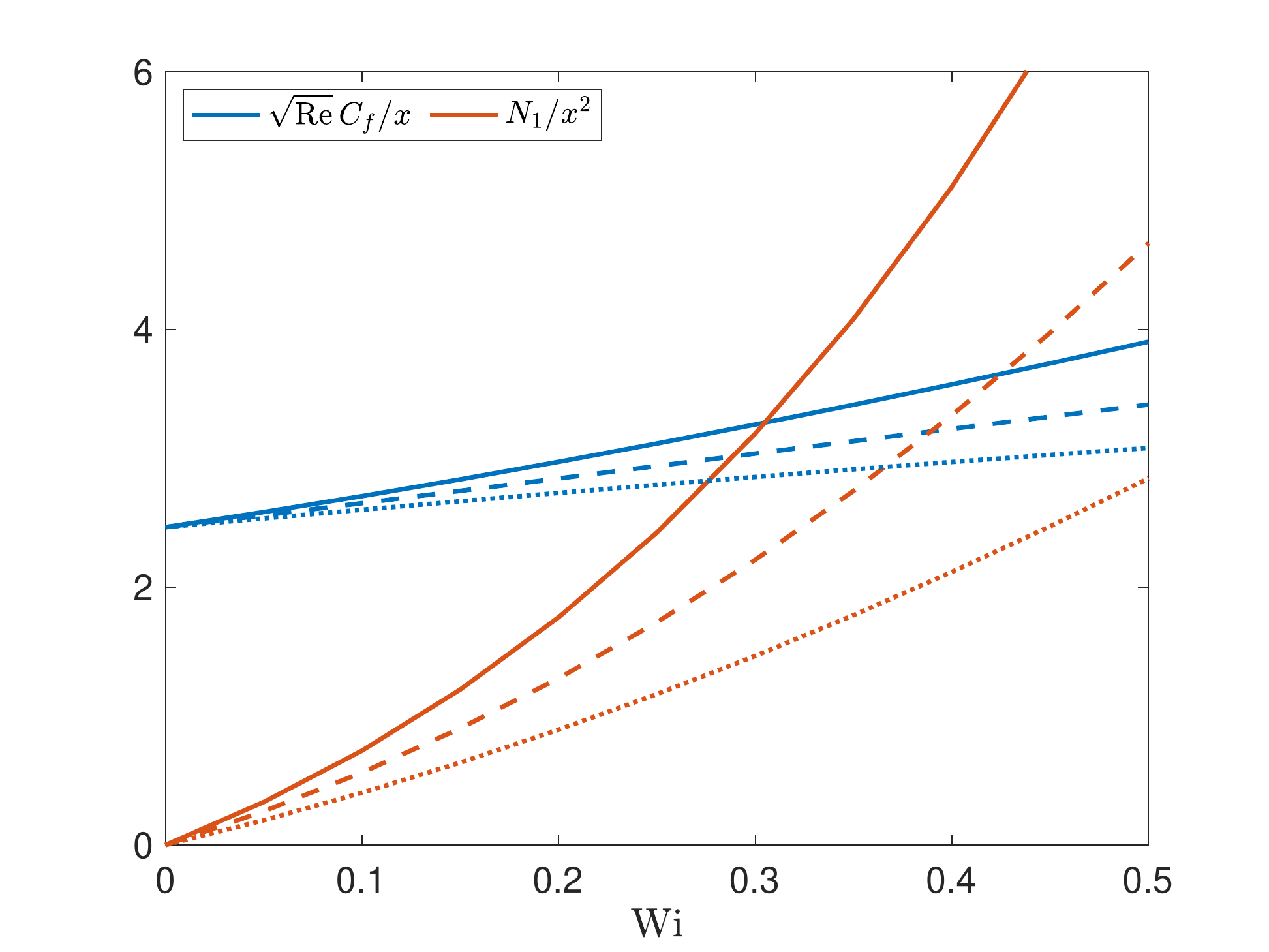}
\caption{Variation of the skin friction coefficient and the first normal stress difference at the wall for three values of the dimensionless retardation parameter $\beta$ and a range of values of the Weissenberg number. The solid curves represent the $\beta=1$ solutions, the dashed curves the $\beta=0.8$ solutions, and the dotted curves the $\beta=0.6$ solutions.}\label{fig:C_f_N_1}
\end{figure}

It is evident, from the results presented in Figure \ref{fig:beta_var_Wi} \textcolor{blue}{(a)}, that as the value of the Weissenberg number is increased, so does the streamwise velocity within the boundary layer. Similarly, as evidenced in Table \ref{bfs} and Figure \ref{fig:C_f_N_1}, the value the skin friction coefficient, and therefore also the first normal stress difference at the wall, increases with increasing $\textrm{Wi}$. This is an indication that the effects of viscoelasticity act to thin the boundary layer profile. Indeed, the fluid particles accelerate near the wall and obtain the value of the free-stream velocity closer to the surface of the plate, primarily because of the increased difference in the first normal stress difference at the wall, with increasing Weissenberg number. This correlates directly to a decreasing of the displacement thickness, $\delta_{1}$, as the value of $\textrm{Wi}$ increases.

From Figure \ref{fig:beta_var_Wi} \textcolor{blue}{(b)}, \textcolor{blue}{(c)}, and \textcolor{blue}{(d)} we are able to correlate our approximate predictions for the decay of stress functions, in the limit as $y\to\infty$, with our numerical solutions. We observe, as predicted by \eqref{decay}, the rapid decay of the streamwise stress function $\tau_{xx}/x^{2}$, when compared to both the shear stress function $\tau_{xy}/x$, and the wall normal stress function $\tau_{yy}$. For unconfined boundary layer flows, such as the flows analysed here, the shear rate is largest at the wall. As one moves away from the wall the shear rate dissipates and the free-stream conditions are attained. The introduction, therefore, of fluid elasticity generates, at the wall, a non-zero stress contribution parallel to the direction of the flow. This explains the form of the $\tau_{xx}/x^{2}$ profiles for increasing values of the Weissenberg number. In a similar fashion, as we transition from analysing a purely viscous fluid ($\textrm{Wi}=0$), to a fluid with elastic properties ($\textrm{Wi}>0$), the shear stress at the wall increases as a result of the introduction of elastic forces, hence the form of the profiles observed in Figure \ref{fig:beta_var_Wi} \textcolor{blue}{(c)}. We note that the boundary conditions imply that the wall normal stress function, $\tau_{yy}$, will always be zero at the wall. From \eqref{tauyy} we see that $\tau_{yy}$ is proportional to $\partial v/\partial y=-f'=-u/x$, it must, therefore, be the case that this function attains a constant negative value at the outer edge of the boundary layer. From Figure \ref{fig:beta_var_Wi} \textcolor{blue}{(d)} we observe that the introduction of elasticity acts to reduce the magnitude of the wall normal stress component at the free-stream. This result correlates, physically, with the ability of the elastic forces to resist the action of the inertia forces directed towards the surface of the flat plate.

We note that, although not plotted here, as the value of $\beta$ decreases from $1$, for a fixed value of the Weissenberg number, the thickness of the boundary layer increases, and the values of both the skin friction coefficient and first normal stress difference at the wall decrease. Physically, these results match with ones' intuition. Decreasing the value of $\beta$ effectively reduces the ability of the fluid to overcome retardation effects and thus, for a fixed free-stream velocity, one would expect to observe a thickening of the boundary layer.

\subsection{Moderate \texorpdfstring{$\textnormal{Wi} \sim O(1)$}{}}\label{subsec:largeWiRes}

The spectral method we use to calculate the flow profiles is not limited to relatively low values of Weissenberg number, as is frequently considered in the literature. We find that, given the number of collocation points as prescribed, the system can be well described up to a value of $\textrm{Wi} \approx 3$. This value falls well within the $O(1)$ restriction which was imposed in the non-dimensionalisation procedure outlined in \S \ref{sec:form}.

\begin{figure}[t!]
\centering
\begin{subfigure}[b]{0.48\textwidth}
\centering
\includegraphics[width=\textwidth]{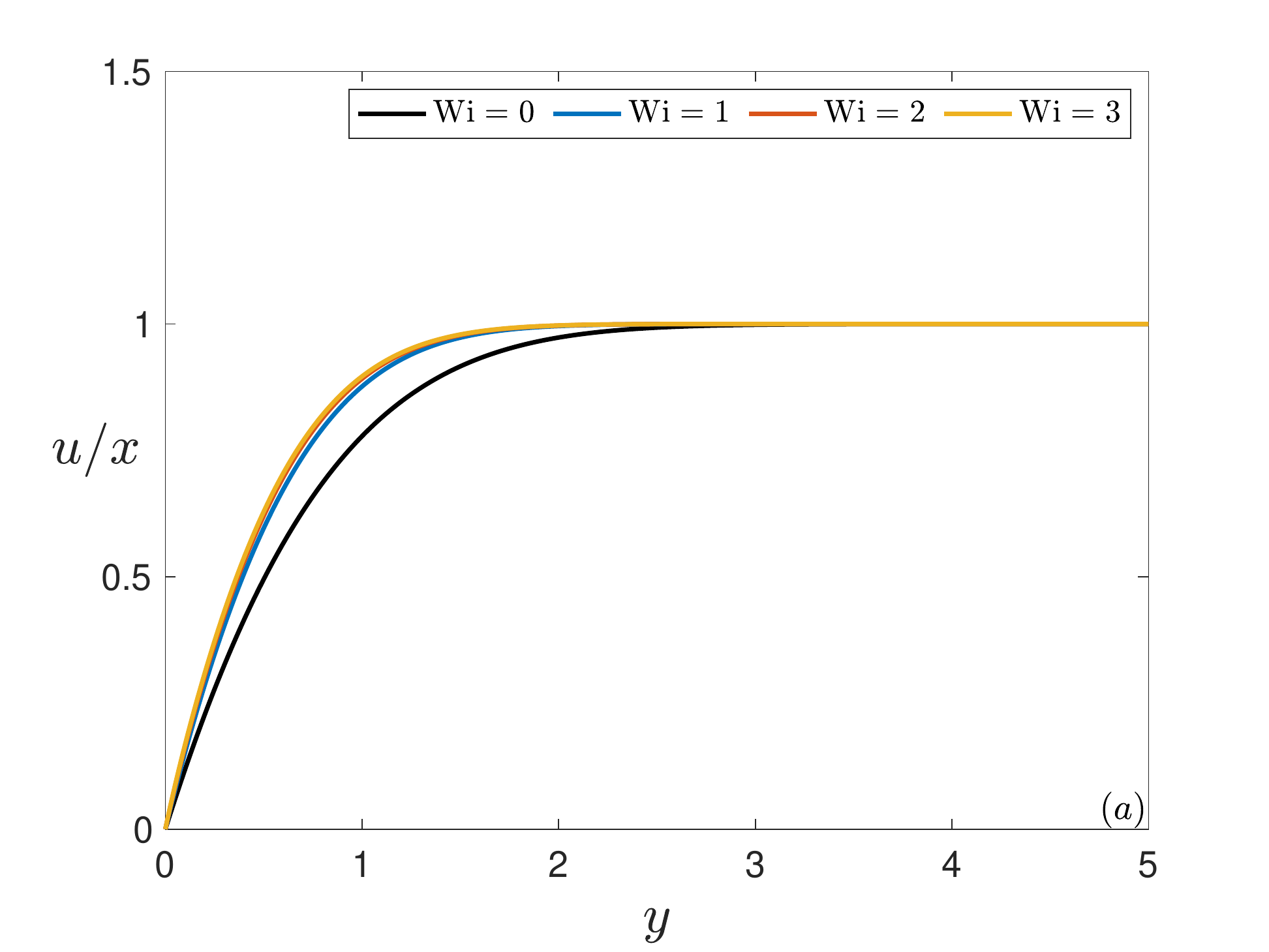}
\label{fig:new_fig_u}
\end{subfigure}
\begin{subfigure}[b]{0.48\textwidth}
\centering
\includegraphics[width=\textwidth]{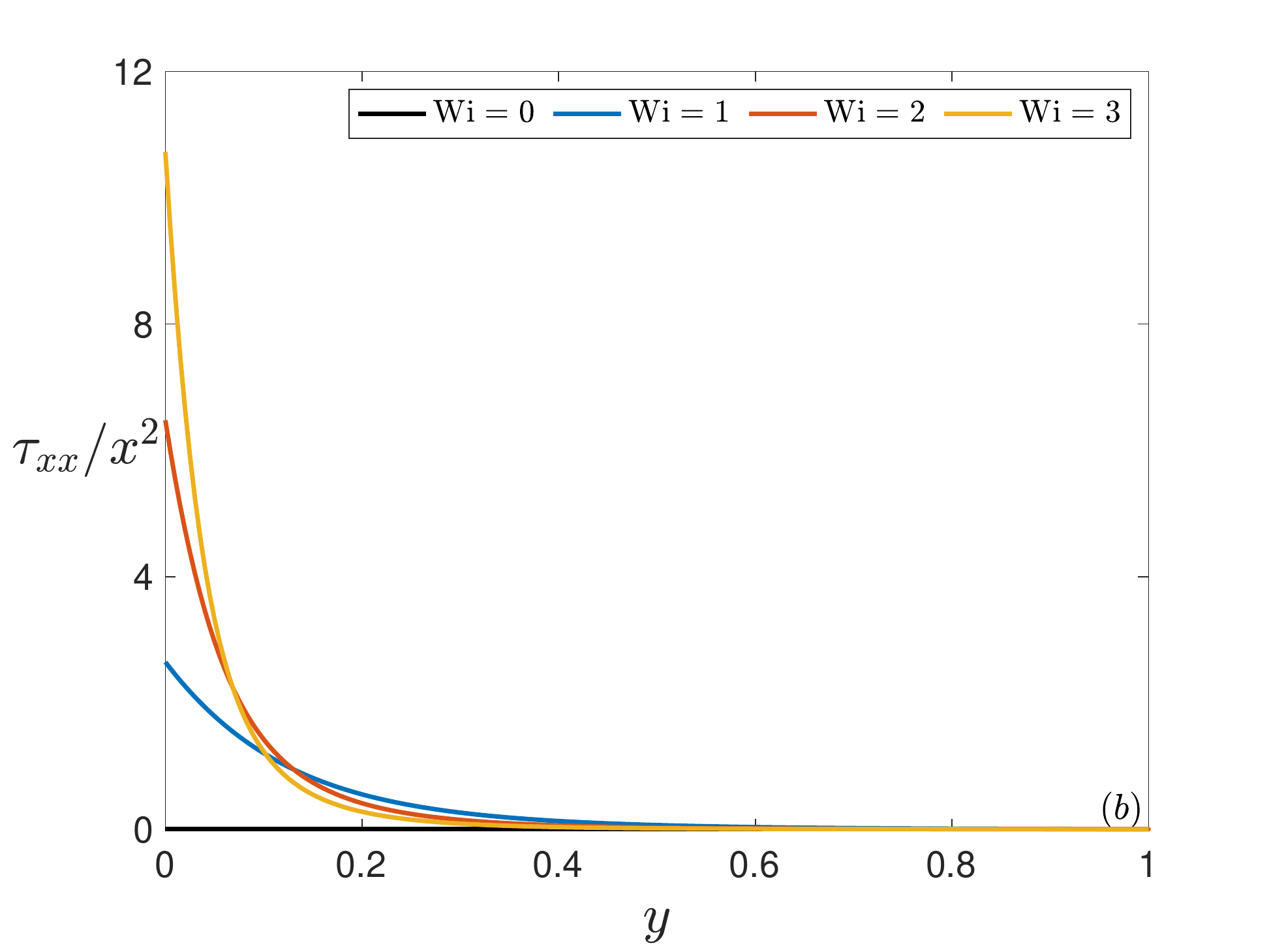}
\label{fig:new_fig_tau_xx}
\end{subfigure}
\begin{subfigure}[b]{0.48\textwidth}
\centering
\includegraphics[width=\textwidth]{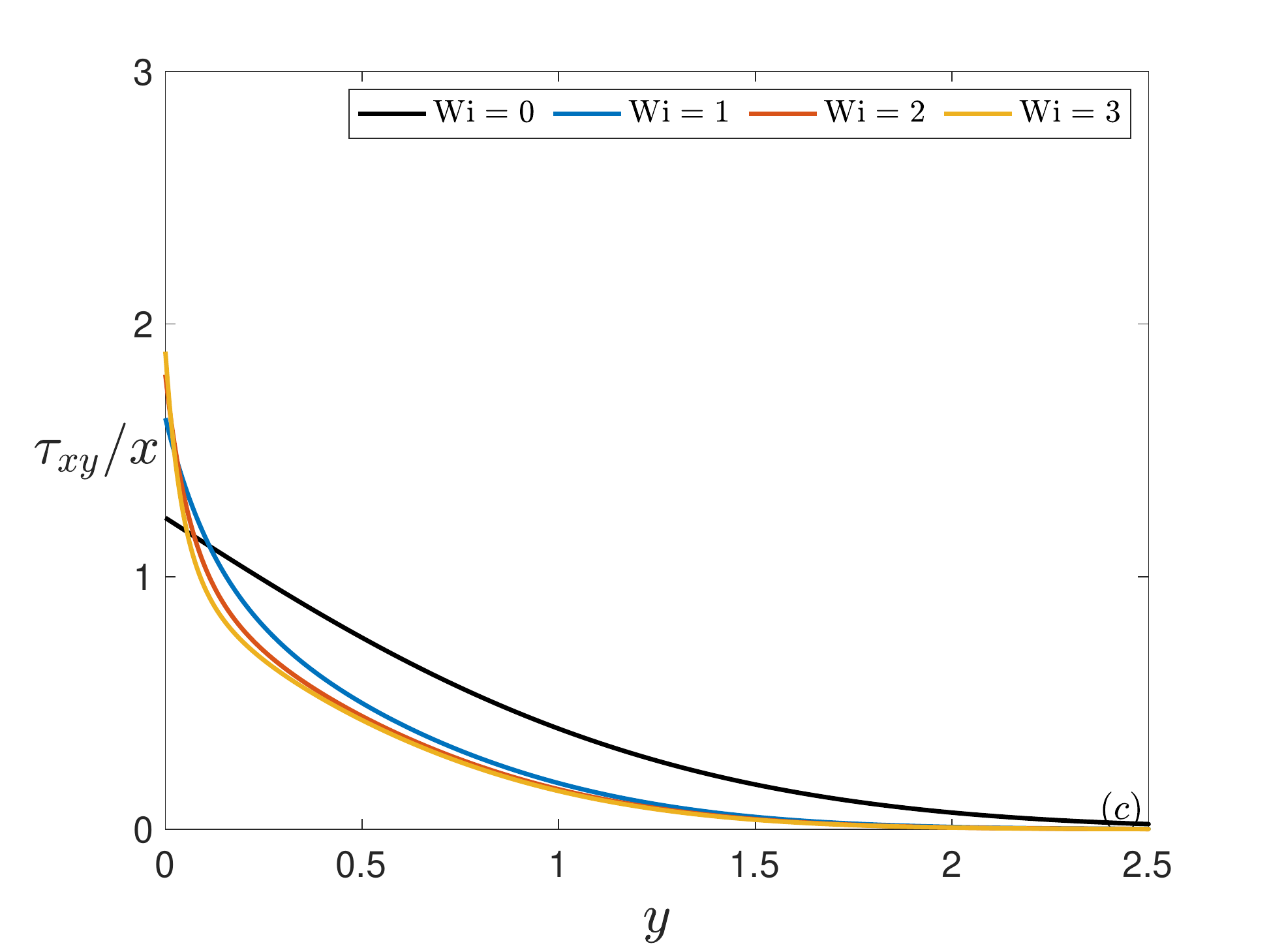}
\label{fig:new_fig_tau_xy}
\end{subfigure}
\begin{subfigure}[b]{0.48\textwidth}
\centering
\includegraphics[width=\textwidth]{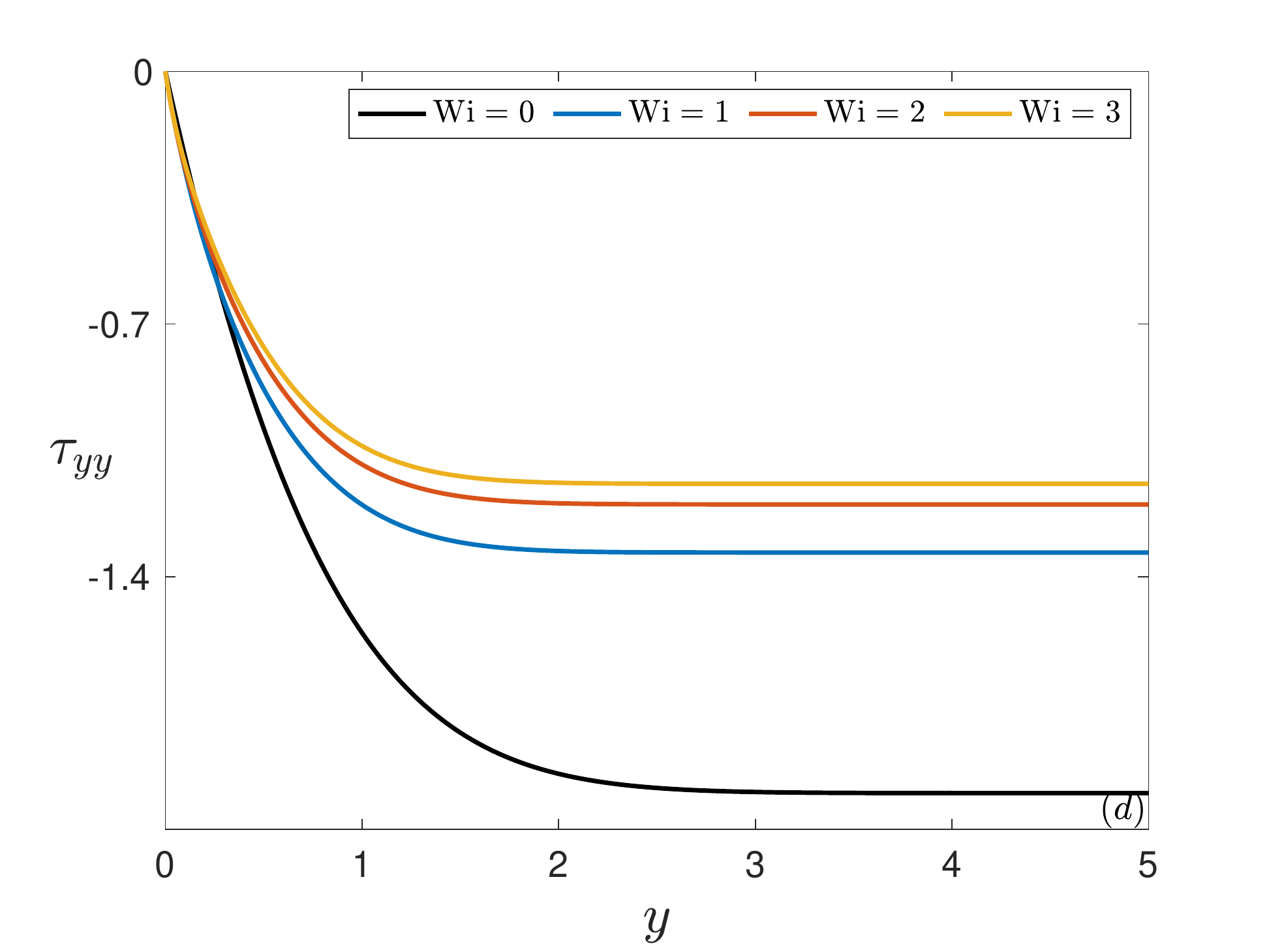}
\label{fig:new_fig_tau_yy}
\end{subfigure}
\caption{In (a) we plot the streamwise velocity $u/x$, against $y$, for a range of moderate values of the Weissenberg number. In (b), (c), and (d) we plot the variation of the three stress functions for the same range of values of the Weissenberg number. In all cases the dimensionless retardation parameter is fixed such that $\beta=0.5$.}\label{fig:new_fig}
\end{figure}

Figure \ref{fig:new_fig} shows the dependence of our horizontal velocity and stress components, given a variety of moderate $O(1)$ Weissenberg numbers and a constant value of the retardation parameter, $\beta = 0.5$. We can see that, similarly to the smaller $\textrm{Wi}$ cases, an increase in the Weissenberg number yields a return to free-stream profiles closer to the wall. As noted previously, an increase in viscoelastic effects within the fluid phase therefore leads to a decrease in the boundary layer thickness.

Figure \ref{fig:new_fig} \textcolor{blue}{(a)} shows a seeming convergence to fixed flow profile as \textrm{Wi} increases. This likely indicates that there is a threshold for \textrm{Wi} at which the viscoelastic contributions to the governing equations become dominant. We note that the dependence is clearly non-linear, and we confirm that intermediate Weissenberg number velocities converge in a monotonic fashion. The investigation into $\textrm{Wi} \sim O(\delta^{-1})$ simulations, given the same scalings provided in \S \ref{sec:form}, is left as an open problem.

Stress functions $\tau_{xx}$, $\tau_{xy}$ and $\tau_{yy}$ are plotted in Figure \ref{fig:new_fig} \textcolor{blue}{(b)}, \textcolor{blue}{(c)}, and \textcolor{blue}{(d)}, respectively. Qualitatively similar results are obtained to those presented in \S~\ref{subsec:smallWiRes}. The main difference between the two sets of results is, however, the magnitude of effects. For example, we observe that the stress function $\tau_{xx} / x^2$ attains a much higher peak at the wall when compared to Figure \ref{fig:beta_var_Wi} \textcolor{blue}{(b)}. Furthermore, we note that the free-stream value in all the stress components is attained closer to the wall, which mimics the behaviour of the streamwise velocity.

\begin{figure}[t!]
\centering
\begin{subfigure}[b]{0.48\textwidth}
\centering
\includegraphics[width=\textwidth]{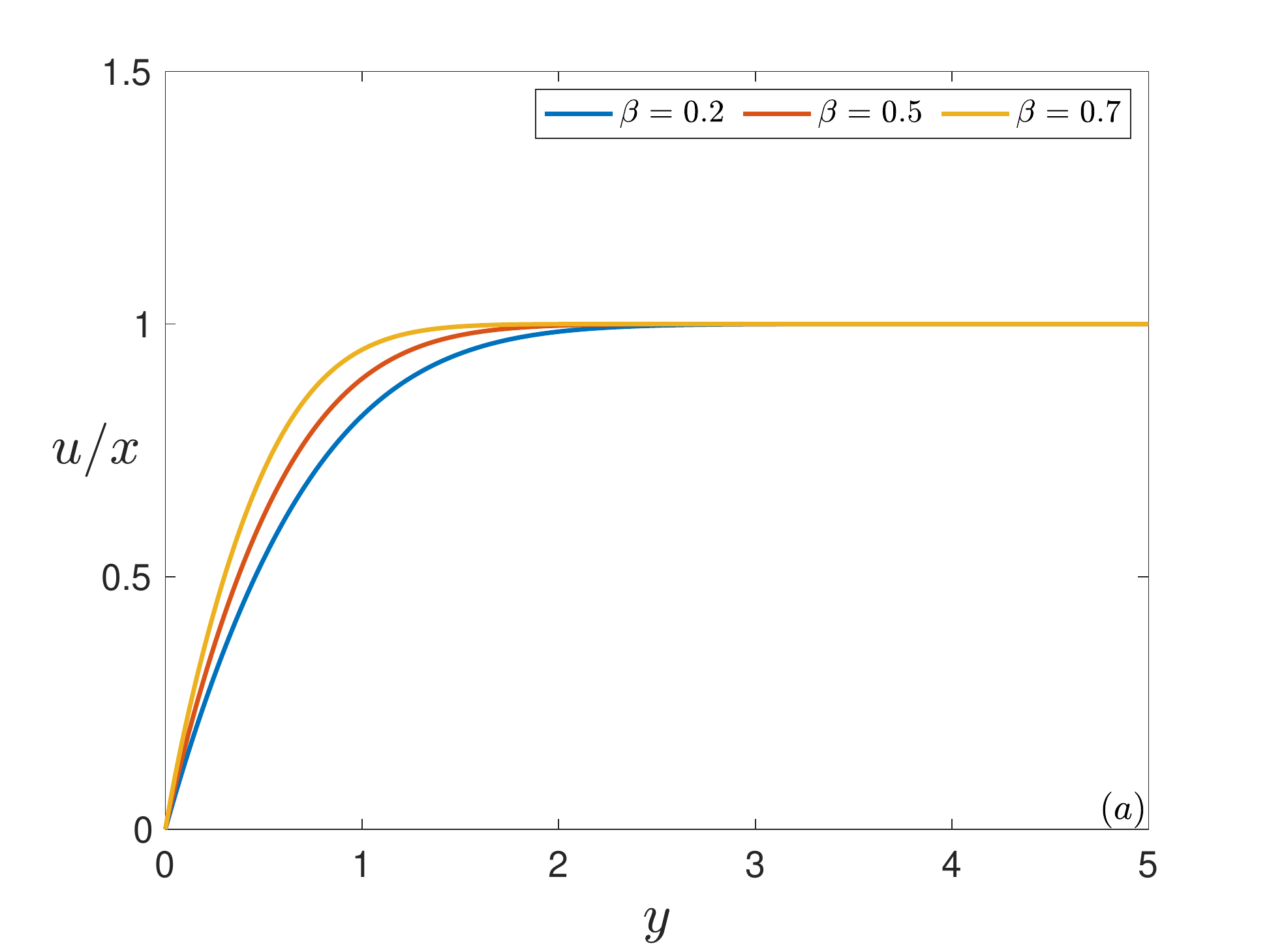}
\label{fig:new_fig_2_u}
\end{subfigure}
\begin{subfigure}[b]{0.48\textwidth}
\centering
\includegraphics[width=\textwidth]{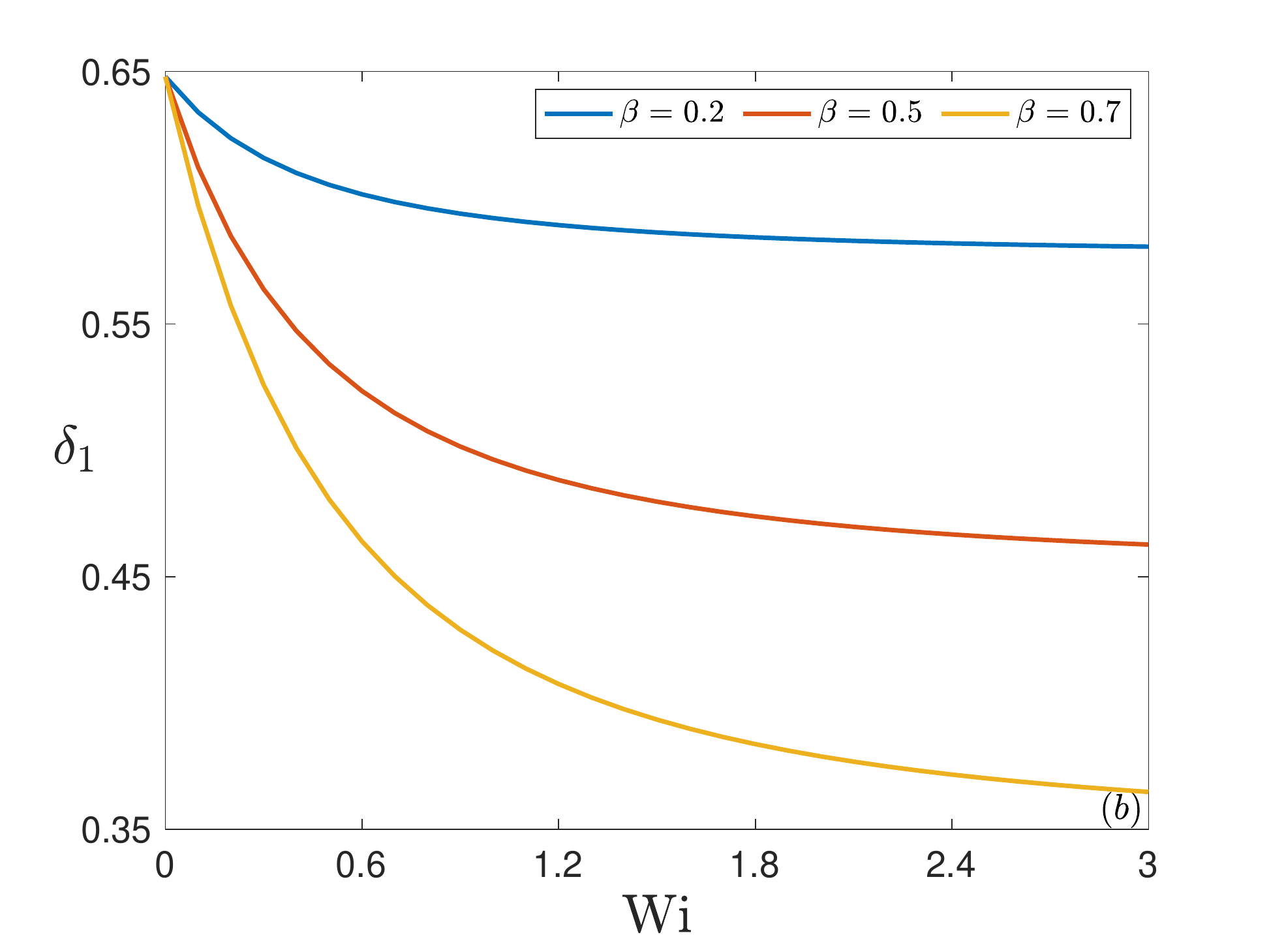}
\label{fig:new_fig_2_delta}
\end{subfigure}
\caption{In (a) we plot the streamwise velocity $u/x$, against $y$, for a range of value of the retardation parameter, given a fixed moderate $\textrm{Wi}=2$. In (b) we plot the variation of the displacement thickness for moderate values of the Weissenberg number, for the same range of $\beta$ values. The Newtonian reference solution $\delta_{1}=0.6479$, is returned in each case when $\textrm{Wi}=0$.}\label{fig:new_fig_2}
\end{figure}

For fixed $\textrm{Wi}=2$, we evaluate the variation of the streamwise velocity profile $u / x$ as the retardation parameter $\beta$ decreases. We observe in Figure~\ref{fig:new_fig_2} \textcolor{blue}{(a)} that as $\beta$ reduces in value, which represents a decrease in polymer viscosity (compared to solvent viscosity), the streamwise velocity decreases accordingly. This results in a thickening of the boundary layer at fixed moderate values of the Weissenberg number.

In Figure~\ref{fig:new_fig_2} \textcolor{blue}{(b)} we plot the variation of the displacement thickness, $\delta_{1}$, for a range of moderate Weissenberg numbers and our familiar choices of the retardation parameter. Naturally, the same behaviour as described above in Figure~\ref{fig:new_fig_2} \textcolor{blue}{(a)} is confirmed here: the boundary layer thickens as the value of $\beta$ decreases. Furthermore, we observe that $\delta_1$ appears to approach a fixed constant value as $\textrm{Wi}$ increases towards the upper end of the spectrum of moderate values. We note that this result is not immediately evident from an analysis of the governing equations. Independent of the value of $\beta$, all curves show a monotonic decreasing nature, which has strongest gradient at the Newtonian limit. They also show clearly that any introduction of viscoelastic effects, or indeed a polymer viscosity will produce a thinning of the boundary layer, when compared to the Newtonian constant value $\delta_1 = 0.6479$. Indeed, when $\beta=0.7$, and $\textrm{Wi}=3$, we observe an approximate halving the thickness of the boundary layer when compared to the Newtonian counterpart.\vspace{\baselineskip}

\section{Comparisons With Previous Studies}\label{sec:comp}

As noted in \S \ref{sec:intro}, numerous previous studies have considered the stagnation point boundary layer flow problem for viscoelastic fluids, with particular emphasis being focused on the upper-convected Maxwell model ($\beta=1$). To the best of our knowledge, all previous studies are confined to the cases when $\textrm{Wi}\le1/2$. For this reason, we restrict comparisons to our small $\textrm{Wi}$ case study. We note that, in contrast to this high Reynolds number study, in the limit of creeping flow ($\textrm{Re}\ll1$), many previous investigation have not been limited to this restrictive range of values of the Weissenberg number and, instead, choose to scale $\textrm{Wi}$ with some inverse power of the boundary layer thickness \citet{Renardy1997}, \citet{Renardy2010}, \citet{Evans}.

\subsection{Upper-convected Maxwell fluid model}\label{subsec:UCMcomp}

In the case when $\beta=1$, the governing ODEs for this problem, \eqref{govODEs}, reduce to those representing an upper-convected Maxwell constitutive viscosity law. In this instance we are able to compare our results with those of \citet{Sadeghyetal2006} who, following the analysis of \citet{Harris}, arrived at the following ODE that is claimed to model the stagnation point flow of a fluid exhibiting an upper-convected Maxwell viscosity
$$f'''+ff''+1-(f')^{2}+k(f^{2}f'''-2ff'f'')=0,$$
where $k$ is referred to as the elasticity number. Given our notation, we will consider this to be equivalent to our definition for the Weissenberg number. The authors solve this ODE subject to the standard boundary layer conditions noted previously
$$f(y=0)=f'(y=0)=0,\quad f'(y\to\infty)\to1.$$
Given the relatively simple form of this third order ODE one could use a variety of different numerical methods to arrive at accurate solutions. Indeed, \citet{Sadeghyetal2006} (from this point onwards referred to as SHT), opted to use a Chebyshev spectral scheme, somewhat similar in nature to our own scheme described in \S\ref{sec:res}. SHT produce results for $\textrm{Wi}$ in the range $\textrm{Wi}=[0,0.3]$ and analyse both the streamwise velocity profile and the displacement thickness. We reproduce their results using a simplified version of our own spectral scheme. Qualitatively, our reproduction of SHT's results appears to be exact. However, we are unable to make a quantitative statement due to the fact that SHT did not report any numerical values from which we could benchmark.

\begin{figure}[t!]
\centering
\begin{subfigure}[b]{0.48\textwidth}
\centering
\includegraphics[width=\textwidth]{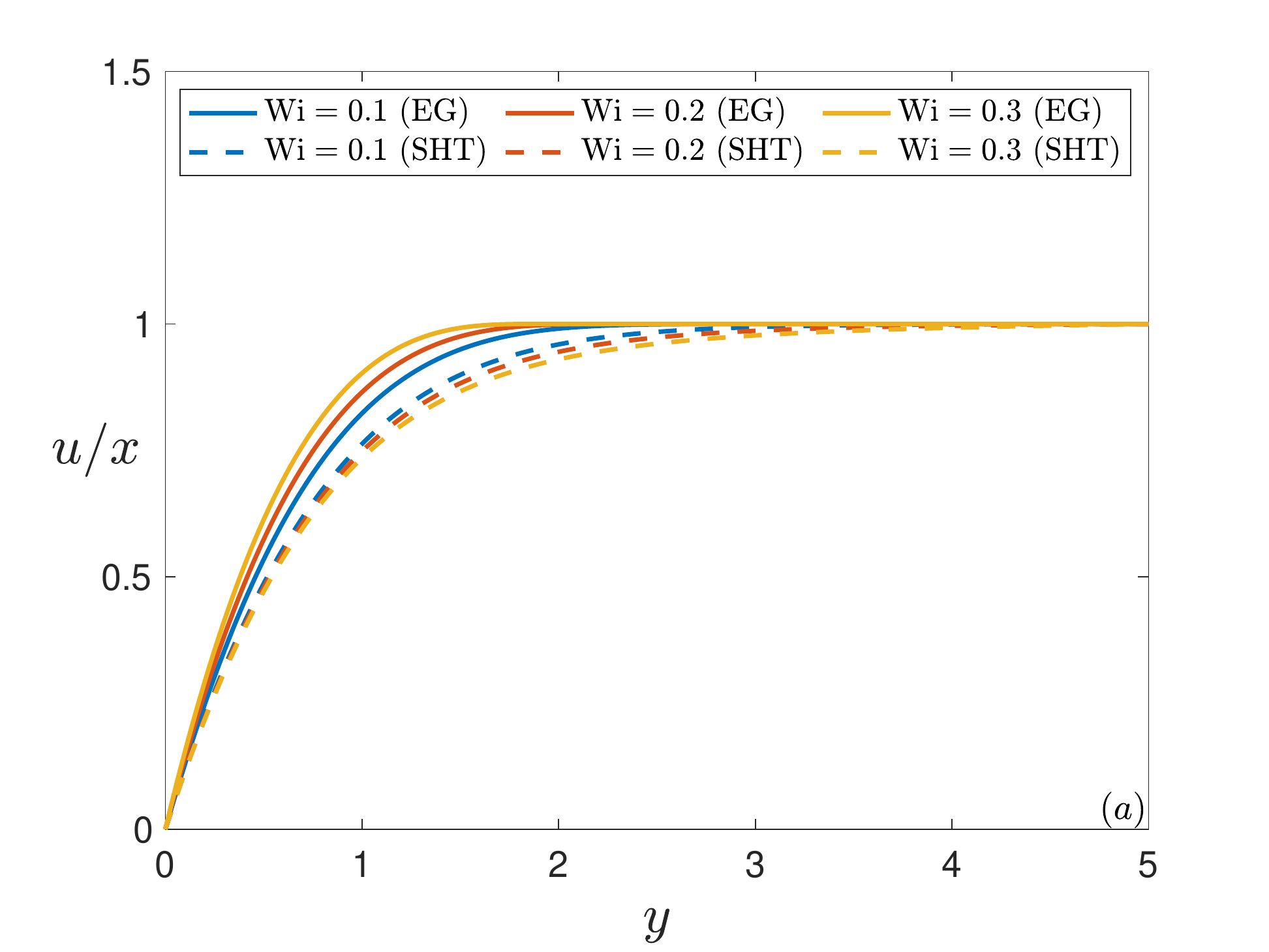}
\label{fig:Wi_var_EG_SHT}
\end{subfigure}
\begin{subfigure}[b]{0.48\textwidth}
\centering
\includegraphics[width=\textwidth]{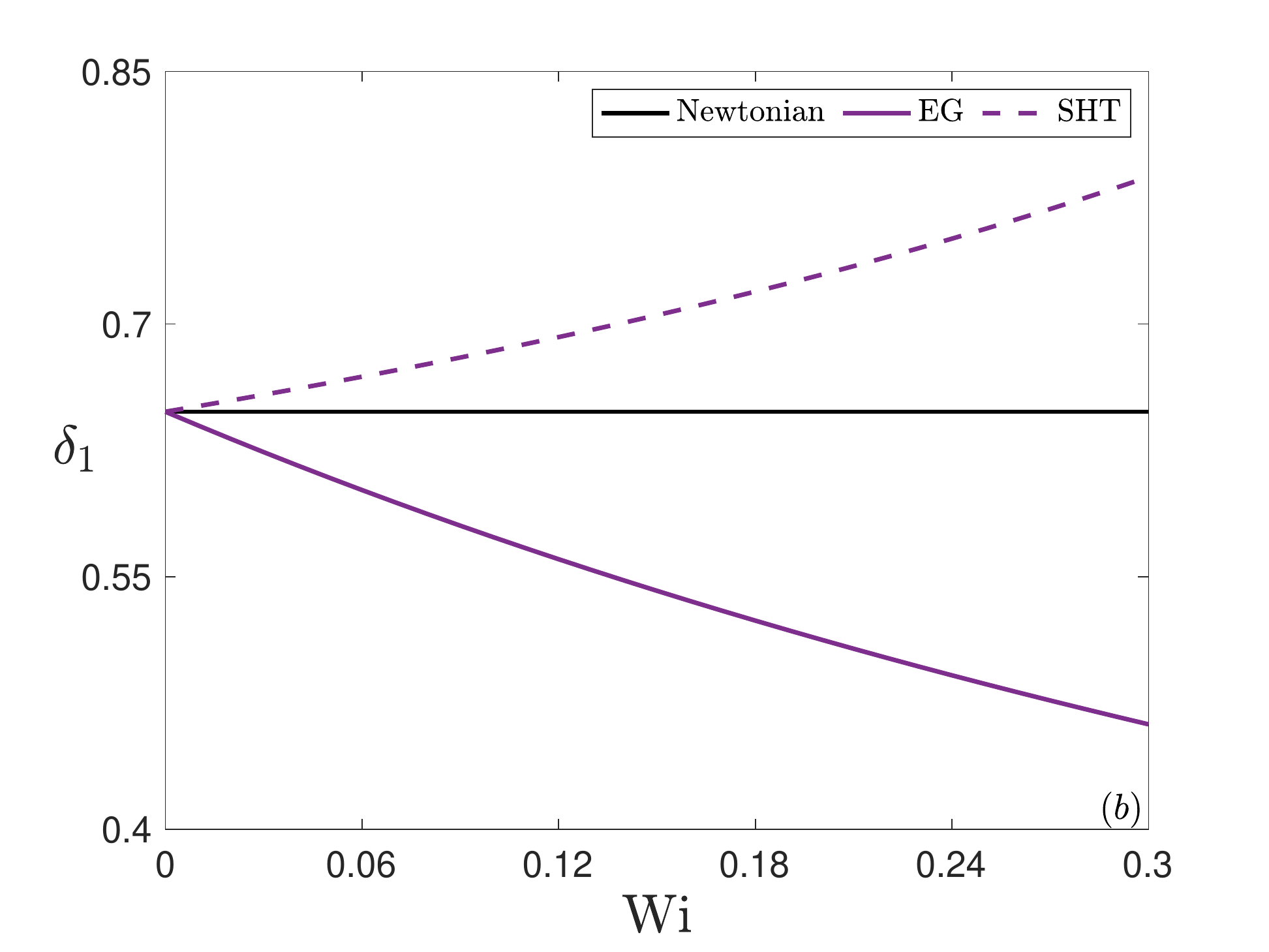}
\label{fig:delta_1}
\end{subfigure}
\caption{In (a) we plot the streamwise velocity $u/x$, against $y$, for a range of values of the Weissenberg number. Our solutions are represented by the solid-line curves whilst those of \citet{Sadeghyetal2006} are represented by the dashed-line curves. In (b) we plot the variation of the displacement thickness with the Weissenberg number. The Newtonian result $\delta_{1}=0.6479$ (black solid-line), is plotted across the full $\textrm{Wi}$ range to serve as a point of reference.}\label{fig:comp1}
\end{figure}

In Figure \ref{fig:comp1} we compare our results with those of SHT for both of these quantities. We observe that as the value of the dimensionless viscoelastic parameter, $\textrm{Wi}$, increases from $0$ (Newtonian flow) our results differ in every respect when compared to those of SHT. As the Weissenberg number increases from zero our results indicate that the free steam velocity will be attained closer to the surface of the flat plate. This result is a direct consequence of the predicted increase of shear stress at the wall (see Table \ref{bfs}). Naturally, this is reflected in a thinning of the boundary layer and a reduction in the value of the displacement thickness $\delta_{1}$. On the other hand, SHT's analysis predicts that the shear stress at the wall will decrease and the boundary layer itself will become broader when compared to its Newtonian counterpart. These results are contrary not only to those owing from our analysis of the full governing equations but also to those of \citet{BeardWalters} who predict, in the limit when $\textrm{Wi}\ll1$, that both the velocity in the boundary layer, and stress at the solid boundary, will increase because of the effect of viscoelasticity. 

\begin{figure}[t!]
\centering
\includegraphics[width=.75\textwidth]{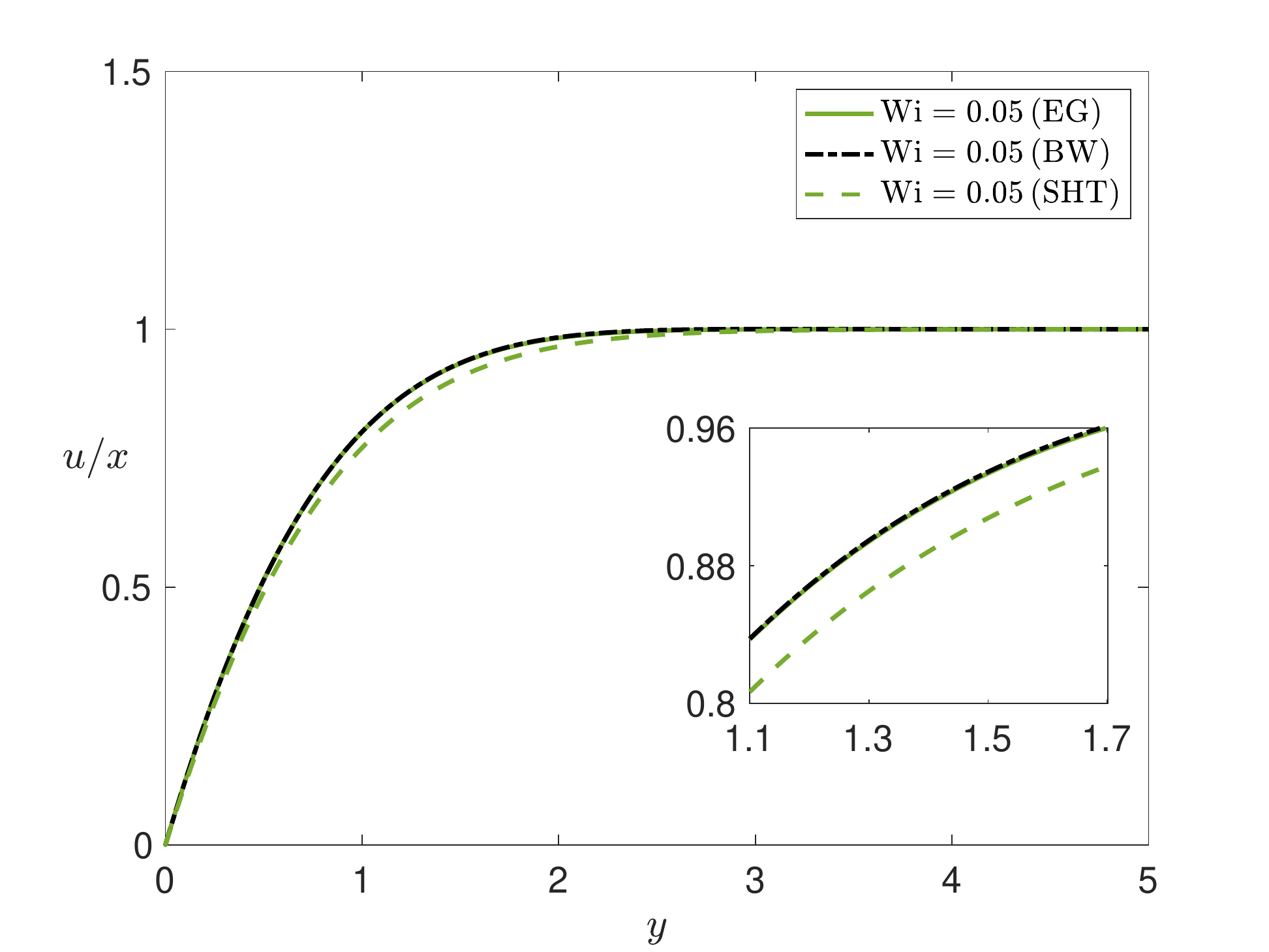}
\caption{Variation of the streamwise velocity $u/x$, against $y$ in the specific case when $\textrm{Wi}=0.05$. Our result is represented by the solid-line curve, that of \citet{BeardWalters} is represented by the dash-dotted black curve, and that of \citet{Sadeghyetal2006} is represented by the dashed-line curve.}\label{fig:EG_BW_SHT_comp}
\end{figure}

In Figure \ref{fig:EG_BW_SHT_comp} we compare our solution for the streamwise velocity component with those of SHT and also \citet{BeardWalters} (from this point onwards referred to as BW). In order to reproduce the results of BW we solve their governing equations ((23) -- (24)) subject to the relevant boundary conditions ((25) -- (26)) using a fourth-order Runge-Kutta (RK4) integrator alongside a Newton-Raphson shooting routine. In the case when $\textrm{Wi}=0.05$, BW predict that $\tau_{xy}(0)=1.2895$. When we attempt to reproduce their analysis we find that $\tau_{xy}(0)=1.2896$. The relative closeness of these results suggest that our RK4 scheme correctly reproduces the results of BW (one would not necessarily expect the two values to match identically given that we are able to produce results to a much higher numerical tolerance). Given that BW's analysis is only strictly valid for small values of the Weissenberg number, and in this case we have set $\textrm{Wi}=0.05$, it is reaffirming that their result match almost identically with our solution obtained from the governing equations~\eqref{govODEs} and boundary conditions~\eqref{govBCs}. In contrast, SHT's result does not closely match that of BW's, something one would expect given the relative smallness of the Weissenberg number.

SHT argue that the difference between their solutions and those of BW are because BW are essentially analysing a second-order fluid model problem. Whereas, they claim to be solving the equations relevant to a UCM boundary-layer flows. However, as we have shown, for flows of this nature, the UCM model collapses down to the second-order fluid model in instances when the Weissenberg number is small. Thus, one would not expect there to be a vast disparity between the results of the two studies for the range of values of the Weissenberg number that were analysed $\textrm{Wi}=[0,0.3]$. Indeed, SHT note that their `\textit{work serves
to demonstrate quite clearly that the constitutive equation of a
fluid may be of crucial importance in boundary layer studies
of viscoelastic fluids}'. However, having followed the incorrect formulation presented by \citet{Harris}, SHT have, in fact, arrived at erroneous conclusions that fail to capture the correct physics of the problem.

Aside from the work of SHT, we can compare our results to those reported by \citet{PhanThien83}. The governing equations used in \cite{PhanThien83} cannot be reduced to a single governing ODE, as they have been in the case of SHT's analysis, and it is a relatively straightforward task to show that Phan-Thien's system of ODEs is directly equivalent to those presented in \S \ref{sec:res}. We note, however, that there is a discrepancy between the boundary conditions, for the three stress functions, that \citet{PhanThien83} employs at the far-field, and our own. Due to a slightly differing formulation to the problem, Phan-Thien's analysis is also restricted to flows whereby the Weissenberg number must satisfy the condition that $\textrm{Wi}<1/2$. The only reported values in that study, to which we can make any comparison, are those for the thickness of the boundary layer. This quantity is defined in \cite{PhanThien83} to be the point at which the streamwise velocity function attains 99\% of the free-stream velocity (from this point onwards we will denote this constant value $\Delta_{99}$). We find that, to the same order of accuracy as given in \cite{PhanThien83} (2 d.p.), the results owing from our spectral approach match precisely with those obtained from Phan-Thien's central difference scheme.

In an attempt to extend the study of SHT, \citet{Mirzadeh2009} return to the stagnation point flow problem, and consider a Giesekus governing viscosity law. This model is only slightly more complex than the upper-convected Maxwell model, and it should be noted that under the correct limiting choice of constants, the Giesekus model reduces in complexity to UCM. However, upon comparison of both our own UCM results (and those of \citet{PhanThien83}) with those presented by \citet{Mirzadeh2009}, we note a discrepancy between the reported values for the boundary layer thickness, $\Delta_{99}$. Again, this is the only tangable reported value that we can make any comparison to. Having analysed the formulation presented in \cite{Mirzadeh2009} with our own (and Phan-Thien's) it would appear that the reason for this discrepancy is due to the relative accuracy of their numerical scheme. \citet{Mirzadeh2009} utilise a Keller box method twinned with a Newton linearisation procedure. Very few details about this linearisation process are forthcoming in the Mirzadeh and Sadeghy's article and it would seem likely that it is this numerical procedure that causes the disparity between the results, as opposed to the aforementioned work of SHT, where clearly the erroneous governing equations are the direct cause of the incorrect solutions.

\subsection{Oldroyd-B fluid model}\label{subsec:OBcomp}

In the more general case when $\beta \neq 1$, the Oldroyd-B fluid equations do not reduce to an explicit upper-convected Maxwell model. We can therefore compare our results to those of~\citet{PhanThien84} (referred to from now on as P--T), who consider a similar, though not identical, stagnation point flow over a flat plate. As previously mentioned in \S \ref{sec:intro}, due to P--T's formulation of the problem he is restricted to considering only a small finite range of the Weissenberg number, $\textrm{Wi} = [0, 0.5]$. Indeed, P--T chooses not to present any results, either tabulated or graphical, for the case when $\textrm{Wi}>0.4$. This is in contrast to our study where no such restriction applies.

\citet{PhanThien84} uses an equivalent set of governing equations as \eqref{govODEs}, the only difference being a reformulation of the governing viscosity equation in terms of $\boldsymbol{\tau}_{NN}^{*} = \boldsymbol{\tau}^{*} - 2 \boldsymbol{E}^{*}$, the non-Newtonian contribution to the stress. This difference in formulation is expected to be of no consequence. However, upon comparing results for the boundary layer thickness $\Delta_{99}$, there is a clear difference between our results and those of P--T. We provide our calculated values in Table \ref{bltcomp} to an accuracy of 5 significant figures, along with the literature values (3 s.f.) as well as the relative difference between the results for the range of $\textrm{Wi}$ and $\beta$ values quoted by P--T.

\begin{table}
\small
\begin{center}
\begin{tabular}{|c|c|c|c|c|c|c|c|c|c|c|c|c|}
\hline
$\textrm{Wi}$ & \multicolumn{3}{|c|}{$\beta=1$} & \multicolumn{3}{|c|}{$\beta=0.7$} & \multicolumn{3}{|c|}{$\beta=0.5$} & \multicolumn{3}{|c|}{$\beta=0.2$} \\
\cline{2-13}
& EG & P--T & $\%$ & EG & P--T & $\%$ & EG & P--T & $\%$ & EG & P--T & $\%$ \\
\hline
$0.1$ & 1.9830 & 1.98 & 0 & 2.1129 & 2.13 & 0.95 & 2.1928 & 2.20 & 0.46 & 2.3127 & 2.33 & 0.87 \\
$0.2$ & 1.6933 & 1.69 & 0 & 1.9281 & 1.95 & 1.04 & 2.0729 & 2.10 & 1.45 & 2.2677 & 2.29 & 0.88 \\
$0.3$ & 1.4735 & 1.47 & 0 & 1.8032 & 1.82 & 1.11 & 1.9880 & 2.00 & 0.50 & 2.2378 & 2.25 & 0.45 \\
$0.4$ & 1.2987 & 1.30 & 0 & 1.7133 & 1.74 & 1.75 & 1.9331 & 1.96 & 1.55 & 2.2128 & 2.23 & 0.90 \\
\hline
\end{tabular}
\caption{A comparison between our results for the boundary layer thickness, $\Delta_{99}$, and those presented by \citet{PhanThien84}. The relative difference between the solutions is noted in each of the third sub-columns.}
\label{bltcomp}
\end{center}
\end{table}

We observe that there is a marked increase in the difference between the results as $\beta$ decreases from its UCM value, $\beta=1$. At this specific value, we match the literature (\cite{PhanThien83} \& \cite{PhanThien84}) precisely to the given accuracy of that paper, which cannot be said in any of the cases when $\beta\neq1$. At its largest, we observed a relative difference between our results and those of P--T of $1.75\%$. We note that we do not see a uniform change in the percentage error for increasing $\textrm{Wi}$ or decreasing $\beta$, as one might expect. This is attributed to the relative accuracy of P--T's stated results. Having said that, in general, we observe that for larger values of both the Weissenberg number and the retardation parameter the error between the two solution sets grows.

The disparity between these two sets of results cannot be due to P--T's differing formulation of governing equations: we have used our in-house spectral code to model exactly P--T's system of ODEs. We find that the values of $\Delta_{99}$ from these calculations match precisely with our own results tabulated in Table \ref{bltcomp}, i.e., when we mimic P--T's analysis we obtain our results. The finite difference scheme employed by P--T is based on a central difference approach, and provides accuracy up to 3 s.f.. However, this cannot account for the source for difference between our results and those of P--T, since even the EG values to the same order of accuracy do match those of P--T. We must conclude, therefore, that there is some small error in the code used by P--T.\vspace{\baselineskip}

\section{Conclusions}\label{sec:conc}

In this paper we have investigated viscoelastic boundary layer flows of fluids described using one of three different models. We have derived the full set of coupled equations in two dimensions which represent all of these fluid models, and further rigorously determined the boundary conditions that must be imposed at the far-field. In the process of doing so, we have highlighted the inadequacies of a number of previous studies in the literature, which fail to capture the correct physics of the problem. The coupled system of governing equations was solved using a Chebyshev collocation method, to a high degree of accuracy.

Results for the streamwise velocity profile and the individual stress components are presented for specific choices of viscoelastic and viscosity parameters. We find that, contrary to the findings of previous investigations, in particular \cite{Sadeghyetal2006}, an increase in viscoelasticity results in a thinning of the boundary layer. Our results tend to agree qualitatively with the work of \citet{PhanThien83}, \citet{PhanThien84} for lower Weissenberg numbers. However, with reference to the determination of the boundary layer thickness, there is a clear disparity between the specific values quoted by these studies and our own. We provide context for the variation of the three stress components across the boundary layer, and show that all of them behave well in the Newtonian limit, where the solution is well known. Furthermore, we evaluate both the skin friction coefficient and first normal stress difference at the wall for fluids with viscosity captured by the Oldroyd-B model. Both of these quantities show a monotonic increase with increasing viscoelastic effects, with the same statement being true as the ratio of the solvent to total viscosity increases.

For moderate values of the Weissenberg number, we present results that retain the high degree of accuracy associated with our spectral method, which would typically be very challenging for finite difference schemes to reproduce. The streamwise flow velocity, along with the stress functions can be seen to approach invariant profiles with increasing values of the Weissenberg number. This behaviour is likely indicative of a transition change from the regime where $\textrm{Wi} \sim O(1)$ to higher orders. We note that at a value of $\textrm{Wi} = 3$, with $\beta = 0.7$, the displacement thickness approximately halves when compared to the Newtonian reference.

Given the relatively general nature of this work, there is a reasonably large range of natural extensions that one may wish to consider. Firstly, the framework presented here could be extended to encapsulate other viscoelastic models. One could consider, for example, extensions to the Phan-Thien Tanner, Giesekus or FENE-P models.

Conversely, one may choose to analyse different types of boundary layer flows whilst considering only the viscoelastic models discussed here. One obvious candidate, given its relative prominence in the literature, would be the boundary layer flow induced by the stretching of a solid surface. In effect, a study of this nature would serve as a correction to the analysis presented by \citet{Sadeghyetal2005}.

At the outset of this study we restricted our attention to moderate Weissenberg number flows. However, there is no reason why the methodology presented here could not be reformatted to solve boundary layer flows associated with large values of the Weissenberg number. Indeed, the governing equations for that problem, as noted by \citet{Evans}, are simply a modified subset of the governing boundary layer equations presented here.

With regards to our original motivation to tackle this problem, we intend to use the results of this study to form the basis for a much wider body of work focusing on the injection of a non-Newtonian fluid in to an otherwise Newtonian boundary layer flow. We expect to be able to report on the results of this study in the relatively near future.

\section*{Acknowledgements}\label{sec:ackn}

The authors wish to acknowledge support from the UKRI EPSRC (Grant No. EP/V006614/2). We also wish to acknowledge the insightful comments of the anonymous reviewers.



\appendix

\section{Boundary layer derivation for quadratic rate of strain term}\label{sec:SOFEEterm}

As noted in \S\ref{sec:form}, the Oldroyd-B fluid model can be described in dimensional terms via the constitutive stress relation~\eqref{stress-gov}, which under the assumption of negligible polymer time scales, can be re-written as
$$\boldsymbol{\tau}^{*}=2\mu_{0}^{*}\boldsymbol{\mathrm{E}}^{*}-2\kappa_{0}^{*}\stackrel{\nabla^{*}}{\boldsymbol{\mathrm{E}}^{*}}.$$
This is a specific case study of the second-order fluid, defined by~\citet{MorozovSpagnolie} as
$$\boldsymbol{\tau}^{*}=2\mu_{0}^{*}\boldsymbol{\mathrm{E}}^{*}+2\alpha_{1}^{*}\stackrel{\nabla^{*}}{\boldsymbol{\mathrm{E}}^{*}}+4\alpha_{2}^{*}(\boldsymbol{\mathrm{E}}^{*}\cdot\boldsymbol{\mathrm{E}}^{*}),$$
with the easiest comparison between the two coming from choice of constants $\alpha_2^{*}=0$, $-\kappa_0^{*}=\alpha_1^{*}$.

The apparent missing dynamics, which originate from the quadratic strain rate tensor term, can be shown to play no role in the solution of velocity or extra stress contributions for flat plate boundary layer flows. We investigate this idea first by assuming a simplified version of the stress tensor:
$$\boldsymbol{\tau}^{*}=2\mu_{0}^{*}\boldsymbol{\mathrm{E}}^{*}+4\alpha_{2}^{*}(\boldsymbol{\mathrm{E}}^{*}\cdot\boldsymbol{\mathrm{E}}^{*}),$$
so chosen to capture the minimum amount of required physics. The first term is the Newtonian contribution, which we know must be dominant in the limit of zero viscoelasticity \cite{Prandtl}, and the second is our non-Newtonian term of interest. This stress profile can be inserted directly into the momentum equations, reducing the number of governing equations from 6 to 3. Under the same non-dimensionalisation and scaling process as outlined in \S\ref{sec:form}, with the definition of a new viscoelastic parameter $k=\alpha_2^{*}U^{*}/\mu_0^{*} L^{*}$, we derive the following set of coupled differential equations:
\begin{align*}
    \partial_{x}u+\partial_{y}v&=0, \\
    u\partial_{x}u+v\partial_{y}u&=-\partial_{x}p+\delta^2\partial_{xx}u+\partial_{yy}u-2k\left(\partial_{y}u\partial_{xy}u+\delta^2\partial_{x}v\partial_{xy}u\right.\\
    & \quad \left.~+4\delta^2\partial_{x}u\partial_{xx}u+\delta^2\partial_{y}u\partial_{xx}v+\delta^4\partial_{x}v\partial_{xx}v\right), \\
    u\partial_{x}v+v\partial_{y}v&=-\delta^{-2}\partial_{y}p+\delta^2\partial_{xx}v+\partial_{yy}v-2k\left(\delta^{-2}\partial_{y}u\partial_{yy}u-4\partial_{x}u\partial_{yy}v\right.\\
    & \quad \left.~+\partial_{x}v\partial_{yy}u-\partial_{y}u\partial_{xx}u-\delta^2\partial_{x}v\partial_{xx}u\right).
\end{align*}
Here we note that the momentum equations above have been simplified via use of the continuity equation.

In the limit as $\textrm{Re}\to\infty$, the above reduces to
\begin{align*}
    \partial_{x}u+\partial_{y}v&=0, \\
    u\partial_{x}u+v\partial_{y}u&=-\partial_{x}p+\partial_{yy}u-2k\partial_{y}u\partial_{xy}u, \\
    0&=-\partial_{y}p-2k\partial_{y}u\partial_{yy}u.
\end{align*}
In a similar fashion to our previous analysis, we solve the $y$ momentum equation for pressure, leading to the succinct formula $p(x,y)=P(x)-k{[\partial_{y}u(x,y)]}^2$, where $P$ is a yet unknown function of $x$. One can see that upon inserting this definition into the $x$ momentum equation, all terms dependent on the viscoelastic parameter $k$ will naturally cancel out:
$$u\partial_{x}u+v\partial_{y}u=-\partial_{x}P+\partial_{yy}u.$$
Since we know $k$ to be a factor of every term in our boundary layer equations which comes from the $\boldsymbol{\mathrm{E}}^{*}\boldsymbol{\cdot}\boldsymbol{\mathrm{E}}^{*}$ contribution, it is clear that both velocities $\boldsymbol{u}^{*}$, and stresses $\boldsymbol{\tau}^{*}$, act independently of it. Instead, we find that the quadratic rate of strain tensor contributes to the pressure function only, in the form of the Newtonian shear derivative squared. We also note that this result is true independent of the angle of inclination of our boundary layer flow.

\section{Decay of the function \texorpdfstring{$\mathcal{T}_{xx}$}{blank} in the far-field}\label{sec:append}

Using the results for the far-field decay of the functions $\mathcal{T}_{xy}$ and $\mathcal{T}_{yy}$, presented in \S\ref{sec:form}, one can then determine the form of the decay of the function $\mathcal{T}_{xx}$, to zero, as $y\to\infty$. From analysis of \textcolor{blue}{(6d)}. The relevant ODE to solve is then
$$Y\mathcal{T}_{xx}'=-2b\beta h^{2}+2a_{1}\beta h^{2}(a_{2}Yh+Y^{2}),$$
where $b=1+a_{1}(2a_{2}-3)$. This differential equation has the decaying solution
$$\mathcal{T}_{xx}\to \beta c^{2}\biggl\{b\, \textrm{e}^{\delta_{1}^{2}}\Gamma(0,Y^{2})-a_{1}\biggl[a_{2} c\,\textrm{e}^{\delta_{1}^{2}}\sqrt{\frac{2\pi \textrm{e}^{\delta_{1}^{2}}}{3}}\textrm{erfc}\biggl(\sqrt{\frac{3}{2}}Y\biggr)-\textrm{e}^{-(Y^{2}-\delta_{1}^{2})}\biggr]\biggr\},$$
where $\Gamma$ is the upper incomplete gamma function, $\textrm{erfc}$ is the complementary error function, and the constant of integration has been chosen such that the free-stream boundary condition is satisfied. Now, to leading order 
$$\Gamma(0,Y^{2})\sim Y^{-2}\textrm{e}^{-Y^{2}},$$
and
$$\sqrt{\frac{2\pi}{3}}\textrm{erfc}\biggl(\sqrt{\frac{3}{2}}Y\biggr)\sim2(3Y)^{-1}\textrm{e}^{-3Y^{2}/2}.$$
Thus the result stated in \S\ref{sec:res} follows
$$\mathcal{T}_{xx}\to a_{1}\beta h^{2}[1-2a_{2}h(3Y)^{-1}+(a_{1}^{-1}+2a_{2}-3)Y^{-2}].$$

\section{Data repository}\label{sec:SpectralData}

An accessible version of our spectral code can be found on GitHub via~\url{https://github.com/L-Escott/OB_Spectral_Code}.



\begin{thebibliography}{10}

\bibitem{Prandtl}
L.~Prandtl.
\newblock {\"{U}ber {F}l\"{u}ssigkeitsbewegung bei sehr kleiner {R}eibung ({O}n the motion of fluids with very little friction)}.
\newblock {\em Verhandl III, Intern. Math. Kongr. Heidelberg, Auch: Gesammelte Abhandlungen}, 2:484, 1905.

\bibitem{Blasius}
H.~Blasius.
\newblock {Grenzschichten in {F}l\"{u}ssigkeiten mit kleiner {R}eibung ({B}oundary layers in fluids with little friction)}.
\newblock {\em Zeitschrift f\"{u}r {M}athematik und {P}hysik}, 56:1, 1908.

\bibitem{Srivastava}
A.~C. Srivastava.
\newblock \href{https://doi.org/10.1007/BF01596862}{The flow of a non-{N}ewtonian liquid near a stagnation point}.
\newblock {\em Z.~Angew.~Math.~Phys.}, 9:80, 1958.

\bibitem{Bhatnagar}
Bhatnagar.
\newblock \href{https://doi.org/10.1007/BF03045795}{On two-dimensional boundary layer in non-{N}ewtonian fluids with constant coefficients of viscosity and cross-viscosity}.
\newblock {\em Ind. Acad. Sci.}, 53:95, 1961.

\bibitem{RajeswariRathna}
G.~K. Rajeswari, and S.~L. Rathna.
\newblock \href{https://doi.org/10.1007/BF01600756}{Flow of a particular class of non-{N}ewtonian visco-elastic and visco-inelastic fluids near a stagnation point}.
\newblock {\em Z. Angew. Math. Phys.}, 13:43, 1962.

\bibitem{RivlinEricksen}
R.~S. Rivlin, and J.~L. Ericksen.
\newblock \href{https://doi.org/10.1007/978-1-4612-2416-7_61}{Stress-deformation relations for isotropic materials}.
\newblock {\em Collected {P}apers of {R}. {S}. {R}ivlin}, 911, 1997.

\bibitem{BeardWalters}
D.~W. Beard, and K.~Walters.
\newblock \href{https://doi.org/10.1017/S0305004100038147}{Elastico-viscous boundary-layer flows {I}. {T}wo-dimensional flow near a stagnation point}.
\newblock {\em Proc. Camb. Phil. Soc.}, 60:667, 1964.

\bibitem{Rajagopaletal1980}
K.~R. Rajagopal, A.~S. Gupta, and A.~S. Wineman.
\newblock \href{https://doi.org/10.1016/0020-7225(80)90035-X}{On a boundary layer theory for non-{N}ewtonian fluids}.
\newblock {\em Int. J. Eng. Sci.}, 18:875, 1980.

\bibitem{Rajagopaletal1983}
K.~R. Rajagopal, A.~S. Gupta, and T.~Y. Na.
\newblock \href{https://doi.org/10.1016/0020-7225(80)90035-X}{A note on the {F}alkner-{S}kan flows of a non-{N}ewtonian fluid}.
\newblock {\em Int. J. Non-Lin. Mech.}, 18:313, 1983.

\bibitem{RajagopalGupta}
K.~R. Rajagopal, and A.~S. Gupta.
\newblock \href{https://doi.org/10.1007/BF01560464}{An exact solution for the flow of a non-{N}ewtonian fluid past an infinite porous plate}.
\newblock {\em Meccanica}, 19:158, 1984.

\bibitem{GargRajagopal1990}
V.~K. Garg, and K.~R. Rajagopal.
\newblock \href{https://doi.org/10.1016/0093-6413(90)90059-L}{Stagnation point flow of a non-{N}ewtonian fluid}.
\newblock {\em Mech Res. Com.}, 17:415, 1990.

\bibitem{GargRajagopal1991}
V.~K. Garg, and K.~R. Rajagopal.
\newblock \href{https://doi.org/10.1007/BF01170596}{Flow of a non-{N}ewtonian fluid past a wedge}.
\newblock {\em Acta Mechanica}, 88:113, 1991.

\bibitem{Ariel}
P.~Ariel.
\newblock \href{https://doi.org/10.1007/BF01170596}{On extra boundary condition in the stagnation point flow of a second grade fluid}.
\newblock {\em Int. J. Eng. Sci.}, 40:145, 2002.

\bibitem{Sadeghyetal2005}
K.~Sadeghy, A.-~H. Najafi, and M.~Saffaripour.
\newblock \href{https://doi.org/10.1016/j.ijnonlinmec.2005.05.006}{Sakiadis flow of an upper-convected {M}axwell fluid}.
\newblock {\em Int. J. Non-Lin. Mech.}, 40:1220, 2005.

\bibitem{Sadeghyetal2006}
K.~Sadeghy, H.~Hajibeygi, and S.-~M. Taghavi.
\newblock \href{https://doi.org/10.1016/j.ijnonlinmec.2006.08.005}{Stagnation-point flow of upper-convected {M}axwell fluids}.
\newblock {\em Int. J. Non-Lin. Mech.}, 41:1241, 2006.

\bibitem{Harris}
J.~Harris.
\newblock {\em Rheology and {N}on-{N}ewtonian {F}low}, pages 249--307.
\newblock Longman, New {Y}ork, 1977.

\bibitem{Hayatetal2006}
T.~Hayat, Z.~Abbas, and M.~Sajid.
\newblock \href{https://doi.org/10.1016/j.physleta.2006.04.117}{Series solution for the upper-convected {M}axwell fluid over a porous stretching plate}.
\newblock {\em Phys. Letters A}, 358:396, 2006.

\bibitem{Hayatetal2011}
T.~Hayat, M.~Awais, M.~Qasim, and A.~A. Hendi.
\newblock \href{https://doi.org/10.1016/j.ijheatmasstransfer.2011.03.003}{Effects of mass transfer on the stagnation point flow of an upper-convected {M}axwell ({UCM}) fluid}.
\newblock {\em Int. J. Heat and Mass Transfer}, 54:3777, 2011.

\bibitem{Abeletal}
M.~S. Abel, J.~V. Tawade, and M.~M. Nandeppanavar.
\newblock \href{https://doi.org/10.1007/s11012-011-9448-7}{MHD flow and heat transfer for the upper-convected {M}axwell fluid over a stretching sheet}.
\newblock {\em Meccanica}, 47:385, 2012.

\bibitem{Mustafaetal}
M.~Mustafa, T.~Hayat, and A.~Alsaedi.
\newblock \href{https://doi.org/10.1016/j.ijheatmasstransfer.2016.10.051}{Rotating flow of {M}axwell fluid with variable thermal conductivity: {A}n application to non-{F}ourier heat flux theory}.
\newblock {\em Int. J. Heat and Mass Transfer}, 106:142, 2017.

\bibitem{PhanThien83}
N.~Phan-Thien.
\newblock \href{https://doi.org/10.1007/BF01332366}{Plane and axi-symmetric stagnation flow of a {M}axwellian fluid}.
\newblock {\em Rheologica Acta}, 22:127, 1983.

\bibitem{PhanThien84}
N.~Phan-Thien.
\newblock \href{https://doi.org/10.1007/BF01332071}{Stagnation flows for the {O}ldroyd-{B} fluid}.
\newblock {\em Rheologica Acta}, 23:172, 1984.

\bibitem{Mirzadeh2009}
M.~Mirzadeh, and K.~Sadeghy.
\newblock \href{https://doi.org/10.1678/rheology.37.31}{On the role played by the extensional behavior of {G}iesekus fluids in plane stagnation flow}.
\newblock {\em J. Soc. Rheo., Japan}, 37:31, 2009.

\bibitem{Olagunju2006a}
D.~O. Olagunju.
\newblock \href{https://doi.org/10.1016/j.aml.2005.05.015}{A self-similar solution for forced convection boundary layer flow of a {FENE}-{P} fluid}.
\newblock {\em Appl. Math. Lett.}, 19:432, 2006.

\bibitem{Olagunju2006b}
D.~O. Olagunju.
\newblock \href{https://doi.org/10.1016/j.amc.2005.04.051}{Local similarity solutions for boundary layer flow of a {FENE}-{P} fluid}.
\newblock {\em Appl. Math. Comput.}, 173:593, 2006.

\bibitem{Parvaretal2021}
S.~Parvar, C.~B. da Silva, and F.~T. Pinho.
\newblock \href{https://doi.org/10.1063/5.0042516}{Revisiting the flat plate laminar boundary layer flow of viscoelastic {FENE}-{P} fluids}.
\newblock {\em Phys. Fluids}, 33:023103, 2021.

\bibitem{MorozovSpagnolie}
A.~Morozov, and S.~E. Spagnolie.
\newblock {\em \href{https://doi.org/10.1007/978-1-4939-2065-5_1}{Introduction to {C}omplex {F}luids}}, pages 3--52.
\newblock Springer, 2015.

\bibitem{KochSubramanian}
D.~L. Koch, and G.~Subramanian.
\newblock \href{https://doi.org/10.1016/j.jnnfm.2006.03.019}{The stress in a dilute suspension of spheres suspended in a second-order fluid subject to a linear velocity field}.
\newblock {\em J. Non-Newt. Fluid Mech.}, 138:87, 2006.

\bibitem{Rallison}
J.~M. Rallison.
\newblock \\href{https://doi.org/10.1017/jfm.2011.544}{The stress in a dilute suspension of liquid spheres in a second-order fluid}.
\newblock {\em J. Fluid Mech.}, 693:500, 2012.

\bibitem{EscottWilson}
L.~J. Escott, and H.~J. Wilson.
\newblock \href{https://doi.org/10.1103/PhysRevFluids.5.083301}{Investigation into the rheology of a solid sphere suspension in second-order fluid using a cell model}.
\newblock {\em Phys. Rev. Fluids}, 5:083301, 2020.

\bibitem{Renardy1997}
M.~Renardy.
\newblock \href{https://doi.org/10.1016/S0377-0257(96)01491-7}{High {W}eissenberg number boundary layers for the upper convected Maxwell fluid}.
\newblock {\em J. Non-Newt. Fluid Mech.}, 68:125, 1997.

\bibitem{Renardy2010}
M.~Renardy.
\newblock \href{https://doi.org/10.1016/j.jnnfm.2009.10.001}{On the high {W}eissenberg number limit of the upper convected Maxwell fluid}.
\newblock {\em J. Non-Newt. Fluid Mech.}, 165:70, 2010.

\bibitem{Evans}
J.~D. Evans.
\newblock \href{https://doi.org/10.1016/j.amc.2019.124952}{High {W}eissenberg number boundary layer structures for UCM fluids}.
\newblock {\em App.~Math.~and Comp.}, 387:124952, 2020.

\end{thebibliography}
\end{document}